\documentclass[a4paper]{article}

\usepackage{epsfig}
\usepackage{subfigure}
\usepackage{calc}
\usepackage{amssymb}
\usepackage{amstext}
\usepackage{amsmath}
\usepackage{multicol}
\usepackage{pslatex}
\usepackage{fancyhdr}
\usepackage[small]{caption}

\usepackage{proof}
\usepackage{ifsym}
\usepackage[colorinlistoftodos, textwidth=4cm, shadow]{todonotes}
\usepackage{color}
\newcommand{\mathii}[1] {\emph{#1}}
\def\mathbi#1{\textbf{\em #1}}

\newcommand{\attentive}[1] {$Attend_{#1}$}
\newcommand{\honest}[1]{$H_{#1}$}
\newcommand{\dishonest}[1]{$A_{#1}$}
\newcommand{\notknow}[1]{$U_{#1}$}
\newcommand{\canHelp}{$\sim$}
\newcommand{\helpAll}{$\surd$}
\newcommand{\puntello}{\rule[-2.5 mm]{0mm}{0.65 cm}}
\newcommand{\puntelloo}{\rule[-2.5 mm]{0mm}{0.58 cm}}

\newcommand{\ofinterest}[1]{$\mathit{ofInterest_{E}(#1)}$}
\newcommand{\cansee}[1]{$\mathit{canSee(#1)}$}
\newcommand{\nethandler}{$\mathit{NetHandler}$}
\newcommand{\trueid}{$\mathit{True\text{-}Sender \text{-}ID}$}
\newcommand{\rspy}{$\mathit{Restricted \text{-}Spy}$}
\newcommand{\ospy}{$\mathit{Outflow\text{-}Spy}$}
\newcommand{\ispy}{$\mathit{Inflow\text{-}Spy}$}
\newcommand{\triple}{$\langle \mathit{sender\textrm{-}ID, message, receiver\textrm{-}ID}\rangle$}

\title{Attack Interference in Non-Collaborative Scenarios for Security
Protocol Analysis [Extended Version]}
\author{
 M.~Camilla Fiazza, Michele Peroli and Luca Vigan\`o
}

\begin{document}

\maketitle

\begin{abstract}
In security protocol analysis, the traditional choice to consider a
single Dolev-Yao attacker is supported by the fact that models with
multiple collaborating Dolev-Yao attackers have been shown to be
reducible to models with one Dolev-Yao attacker. 
In this paper, we take a fundamentally different approach and investigate the case of multiple non-collaborating attackers. After formalizing the framework for multi-attacker scenarios, we show with a case study that concurrent competitive attacks can interfere with each other. We then present a new strategy to defend security protocols, based on active exploitation of attack interference. The paper can be seen as providing two proof-of-concept results: (i) it is possible to exploit interference to mitigate protocol vulnerabilities, thus providing a form of protection to protocols; (ii) the search for defense strategies requires scenarios with at least two attackers.
\end{abstract}

\section{Introduction}

\subsection{Context and motivations}
The typical attacker model adopted in security protocol analysis is the
one of~\cite{MR712376}: the \emph{Dolev-Yao (DY) attacker} can compose,
send and intercept messages at will, but, following the perfect
cryptography assumption, he cannot break cryptography. The DY attacker
is thus in complete control of the network --- in fact, he is often
formalized as being the network itself --- and, 
with respect to network abilities, he is actually stronger than
any attacker that can be implemented in real-life situations.
Hence, if a protocol is proved to be secure under the DY attacker, it
will also withstand attacks carried out by less powerful attackers;
aside from deviations from the specification introduced in the implementation phase, the protocol can thus be safely employed in real-life networks, at least in principle.

Alternative attacker models have also been considered. On the one hand,
\emph{computational models} for protocol analysis consider attackers who
can indeed break cryptography, as opposed to the \emph{symbolic models}
where cryptography is perfect (as we will assume in this paper). See,
for instance, \cite{AbadiBlanchetComonCAV09} for a survey of models and
proofs of protocol security, and~\cite{basincremers_models} for a
protocol-security hierarchy in which protocols are classified by their
relative strength against different forms of attacker compromise.

On the other hand, different symbolic models have been recently proposed
that consider \emph{multiple attackers} instead of following the usual
practice to consider a single DY attacker,
  ~a choice that is supported by the fact that models with multiple collaborating DY
attackers have been shown to be reducible to models with one DY attacker
(see, e.g., \cite{Caleiro200567} for a detailed proof, as well
as~\cite{DTL,comon03twoAgentsEnough,Syverson00dolev-yaois} for general
results on the reduction of the number of agents to be considered). For
instance, \cite{basin_physical,schaller_physical} extend the DY model to
account for network topology, transmission delays, and node positions in
the analysis of real-world security protocols, in particular for
wireless networks. This results in a distributed attacker, or actually
multiple distributed attackers, with restricted, but more realistic,
communication capabilities than those of the standard DY attacker.

Multiple attackers are also considered in the models
of~\cite{arspawits2009GA,ArsacBellaChantryCompagna,Bella03,retaliation},
where each protocol participant is allowed to behave maliciously and
intercept and forge messages. In fact, each agent may behave as a DY
attacker, without colluding nor sharing knowledge with anyone else. The
analysis of security protocols under this multi-attacker model allows
one to consider scenarios of agents competing with each other for
personal profit. Agents in this model may also carry out
\emph{retaliation attacks}, where an attack is followed by a
counterattack, and \emph{anticipation attacks}, where an agent's attack
is anticipated, before its termination, by another attack by some other
agent.

The features of the models of~\cite{basin_physical,schaller_physical}
and
of~\cite{arspawits2009GA,ArsacBellaChantryCompagna,Bella03,retaliation}
rule out the applicability of the $n$-to-1 reducibility result for the
DY attacker, as the attackers do not necessarily collaborate, and might
actually possess different knowledge to launch their attacks. They might
even attack each other. In fact, retaliation and anticipation allow
protocols to cope with their own vulnerabilities, rather than
eradicating them. This is possible because agents are capable of doing
more than just executing the steps prescribed by a protocol: they can
decide to anticipate an attack, or to counter-attack by acting even
after the end of a protocol run (in which they have been attacked).
Still, retaliation may nevertheless be too weak as honest agents can
retaliate only \emph{after} an attack has succeeded, and cannot defend
the protocol during the attack itself.

\subsection{Contributions}

In this paper, we take a fundamentally different approach: we show that
multiple non-collaborating DY attackers may interfere with each other in
such a manner that it is possible to exploit interference to mitigate
protocol vulnerabilities, thus providing a form of protection to 
flawed protocols.

To investigate the non-cooperation between attackers, we propose a
(protocol-inde\-pendent) model in which: (i) a protocol is run in the
presence of multiple attackers, and (ii) attackers potentially have
different capabilities, different knowledge and can interfere with each
other. This, ultimately, allows us to create a 
benign attacker for the
system defense: agents can rely on a \emph{network guardian}, an ad-hoc
agent whose task is diminishing the frequency with which dishonest
agents can succeed in attacking vulnerable protocols. This methodology
moves the focus from an attack-based view of security to a defense-based
view.

In other words, in the approach we propose, instead of looking for
attacks and reacting to the existence of one by redesigning the
vulnerable protocol, we look for strategies for defending against
existing known attacks. We would be performing protocol analysis to identify
possible \emph{defenses}, rather than attacks.

We proceed as follows. In Section~\ref{sec:model}, we formalize models
for the network and the agents, including, in particular, agent
attitude, goals, and disposition. We then consider in
Section~\ref{sec:case-study} a vulnerable 
protocol from~\cite{boydMathuria} as
a case study and focus on the interactions between attack procedures
that cannot be observed in classical settings. In
Section~\ref{sec:defense}, we explain how interference between attacks
leads to a methodology that can be used for defending weak
(vulnerable) protocols
against attacks. In Section~\ref{sec:conclusions}, we conclude by
discussing our approach and current and future work. 
Appendix A provides additional details about the case study; a second case study is explored in appendix B.

\section{System models: network, agents, attitude}
\label{sec:model}

\subsection{Goals of modeling and approach}

Network models for security protocol analysis typically either replace
the communication channel with a single attacker or build dedicated
channels for each attacker
(e.g.~\cite{DTL,Caleiro200567,DillowayLoweWITS07,KamilLowe10,Syverson00dolev-yaois}).
Traditional modeling strategies are not adequate to describe the
non-collaborative scenario under consideration. The main shortcoming is
the fact that the ability to spy the communication on a particular
channel is hard-wired in the network model and may depend critically on
network topology or attacker identity; the result is that an
information-sharing mechanism (or a partial prohibition for it) is
structurally encoded in the network. We would like, instead, to (i)
abstract from positional advantages and focus solely on how attackers 
interfere \emph{by attacking}; (ii) treat information-sharing (also as a 
result of spying) as a strategic choice of the agents.

For simplicity, in this paper we restrict our attention to \emph{two}
non-collaborative attackers ($E_{1}$ and $E_{2}$), in addition to the
two honest agents $A$ and $B$ and a trusted third-party server $S$,
whose presence is required by the protocol under consideration. 
In the following, let \mathii{Eves}$=\{E_1, E_2\}$ be the \emph{set of attackers} and
\mathii{Agents}$=\{A, B, E_1, E_2\}$ the \emph{set of all network agents} (honest and
dishonest, server excluded). Let $X$, $Y$, $Z$ and $W$ be variables varying in \mathii{Agents} and $E$ a variable in \mathii{Eves}; $j$ takes value in $\{1,2\}$, whereas $i \in \mathbb{N}$ is reserved for indexing states.

We are aware that, in situations with more than two (dis)honest agents,
further types of interactions can arise; however, a full comprehension
of the interactions depends on building a clear picture of interference.
Such a picture necessarily starts with the elementary interaction
between two attackers.

In order to focus on the raw interference between two attackers, both
directing their attack towards the same target, it is important for all
attackers to have access to the same view of what is taking place with
honest agents and possibly different views of what is taking place with
the other attacker(s). If attackers do not all have the same
information, it is possible to conceive of strategies in which some
attackers can be mislead by others on purpose.

If the knowledge\footnote{Note that we do not attach any epistemic
interpretation to the knowledge we consider in this paper
we simply consider the information initially available to the agents, together with
the information they acquire during protocol executions.}
available to an
attacker affects his view of the system, attacker capabilities and
effectiveness can be diversified, without needing to construct
asymmetric attackers or hardwire constraints that may hold for some
attackers and not for others. We find it relevant that a network model
for non-collaborative scenarios --- besides reflecting this stance ---
also support a form of competition for access to messages, especially if
attacks rely on erasing messages.

If it is possible in principle to actively interfere with an attack, it
should be possible to do so even if all attackers have the same
knowledge. However, differentiating attackers with respect to their
understanding of the situation --- in particular with respect to
awareness of other attackers --- may bring into focus the conditions, if
any, that allow an attacker to interfere with another without being
interfered with.

We diversify the activity of our attackers by admitting that attackers
may choose to selectively ignore some messages, on the basis of the 
sender's and receiver's identifiers. This choice reflects actual situations
in which attackers pay attention to only a subset of the traffic through
a network, focusing on the activity of some agents of interest.
Regardless of whether this selection is caused by computational
constraints or by actual interest, real attackers filter messages on the
basis of the sender's or receiver's identity. In the following, we will
use the set \attentive{E} to model the agents to which attacker $E$ is
attentive; the predicate \ofinterest{X} (see Table~\ref{OurDYmodel})
models the decisional process of
attacker $E$ as he considers whether he wishes to augment \attentive{E}
with $X$, i.e. \ofinterest{X} implies that $X$ is added into
\attentive{E}.

Honest agents are interested in \emph{security properties} (such as
authentication or secrecy) being upheld through the use of protocols.
Dishonest agents, on the other hand, are interested in changing or
negating such properties.

The characteristic feature of the attackers we consider is their
attitude. In particular, in the case study that we consider in the next
section, dishonest agents wish to attack the security protocol and are
ready, should they encounter unforeseen interference, to take
countermeasures with respect to the interference as well. In a sense,
each attacker is exclusively focused on attacking the protocol and
becomes aware of other attackers through their effect on his success.

Our target is capturing the behavior of \emph{equal-opportunity}
dishonest agents that do not cooperate in the classical sense. By
equal-opportunity attackers we mean agents that have the same attack
power and that differ with respect to the information content of their
knowledge bases. Such differentiation arises out of attentional choices
and not out of intrinsic constraints. Strategic and attitude
considerations should not be derivable explicitly from the attacker
model --- rather, they should configure it.

The driving hypothesis of our work is that studying non-collaboration
requires a complex notion of attacker, whose full specification involves
attentional choices, decisional processes pertaining to the network
environment and to other agents, cooperation-related choices and
decisional processes pertaining to the attack strategy. To support this
type of attacker, we extend the usual notions of protocol and role by
introducing a control --- a mechanism to regulate the execution of the
steps prescribed by the attack trace in accordance with the attacker's
strategy. 
In our model, honest agents perform a controlled execution of the
protocol as well, so as to support in-protocol detection of attacks.
Honest agents behave according to the protocol's prescription, expect
things to go exactly in accordance with the protocol and interpret
deviations in terms of the activity of dishonest agents.

\begin{table}[!t]
\begin{center}
\scalebox{0.84}{
\begin{scriptsize}
\begin{tabular*}{0.555\textwidth}{c}
\hline
\hline \\
$\infer[\mathbi{(Comp)}]{(m_{1}, m_{2})\in D^{i}_{E}}{m_{1}\in D^{i}_{E} \quad m_{2}\in D^{i}_{E}}$
\qquad
$\infer[\mathbi{(Encr)}]{\{m\}_{k}\in D^{i}_{E}}{m\in D^{i}_{E} \quad k\in D^{i}_{E}}$
\\\\
$\infer[\mathbi{(Proj)} \quad]{m_{j}\in D^{i}_{E}  \:\: \text{for~} j \in \{1,2\}}{(m_{1}, m_{2})\in D^{i}_{E}}$ 

$\infer[\mathbi{(Decr)} \quad]{m \in D^{i}_{E}}{\{m\}_{k}\in D^{i}_{E} \quad k^{-1} \in D^{i}_{E}}$
\\\\
%
\hline
\hline \\
$\infer[\mathbi{(Restricted-Spy)}]{m \in D^{i+1}_{E}}{<X, m, Y> \in D_{net}^{i} \quad \text{sender}(<X,m,Y>) \in D^{i}_{E} \quad Y \in D^{i}_{E}\quad \psi}$ 
\\\\
$\infer[(\text{\ispy})]{m \in D^{i+1}_{E} \land sender(<X,m,Y>) \in D^{i+1}_{E} }{<X, m, Y> \in D_{net}^{i} \quad $ \ofinterest{X}$ \quad Y \in D^{i}_{E} \quad \psi} $
\\\\
$\infer[(\text{\ospy})]{m \in D^{i+1}_{E} \land Y \in D^{i+1}_{E}}{<X, m, Y> \in D_{net}^{i} \quad {sender}(<X,m,Y>) \in D^{i}_{E} \quad $\ofinterest{Y}$ \quad \psi }$ 
\\\\
{where  $\psi = E \in $\cansee{<X,m,Y>, i}  }\\\\
\hline
\\
$\infer[\mathbi{(Injection)}]{<E(X), m, Y>  \in D_{net}^{i+1}}{m \in D^{i}_{E} \quad X \in D^{i}_{E} \quad Y \in D^{i}_{E}}$ 
 \\\\
$\infer[\mathbi{(Erase)}]{<X, m, Y> \notin D_{net}^{i+1}}{<X,m,Y> \in D_{net}^{i} \quad {sender}(<X,m,Y>) \in D^{i}_{E}}$
\\\\
\hline \hline
\\
$
{sender}(<X, m, Y>)= \begin{cases}
 E & \text{~if~ there exists $Z$ such that~} X = E(Z) \\
 X & \text{~otherwise}
\end{cases}
$  
{\scriptsize \mathbi{(True-sender-ID)}} \\\\
\ofinterest{X}
$
=  
\begin{cases}
 true & \text{if $E$ decides to pay attention to $X$} \\
 false & \text{~otherwise}
\end{cases}
$
{\scriptsize $(DecisionalProcess)$} \\\\
\hline\\
$
\text{\cansee{<X,m,Y>,i}} \: = \:
 \{Z \in Eves \:|\:   \text{ $Z$ can spy}
 <X,m,Y> \text{~on $D_{net}^{i}$} \}
$ {\scriptsize \mathbi{(NetHandler)}} 
\\\\
\hline
\hline
\end{tabular*}
\end{scriptsize}
}
\caption[Attacker model for non-collaboration]{Dolev-Yao attacker model for non-collaborative scenarios: internal operations (synthesis and analysis of messages), network operations (\mathii{spy}, \mathii{inject}, \mathii{erase}) and system configuration (\trueid{}, \mathii{DecisionalProcess}, \nethandler{}). \nethandler{} describes the set of attackers who are allowed to spy by applying one of the \mathii{spy} rules. We omit the usual rules for conjunction.
The rules employed in the case study are marked in boldface.
}
\label{OurDYmodel}
\end{center}
\end{table}

\subsection{Agent model} \label{attacker}

Agent knowledge is characterized in terms of a proprietary dataset.
To each $X$ in \mathii{Agents}, we associate the dataset $D_X$, which we assume to be monotonically non-decreasing. Our agents, in particular dishonest agents, collect information but do not forget it. When it is important to highlight that the dataset is to be considered at a particular moment, we will use $D_X^i$ instead.
 
The network \mathii{net} is also formalized through a dataset, which is named $D_{net}$ and indexed in the same manner as $D_X^i$. A dataset is a simple network model that can be configured to support complex attackers; we believe it can successfully meet all of our modeling requirements for non-collaboration. We postpone to Section \ref{sec:evolution} the discussion of how datasets evolve and how indexing and evolution are related to actions and message transmission.

We adapt the notion of DY attacker ~\cite{MR712376} to capture a 
non-collaborative scenario. We show in Table~\ref{OurDYmodel} how one 
such attacker is formalized within our model, writing rules 
for attacker $E$ with respect to the knowledge base $D_{E}$ and the network model
$D_{net}$. 
Let us specify that the rules in Table~\ref{OurDYmodel} are transition rules, rather than deduction rules. Taken altogether, they construct a \emph{transition system} -- which describes a computation by describing the states that are upheld as a result of the transition. We do not intend to carry out in this paper logical inference to identify defenses against attacks; rather, we recognize in the system's evolution what in our eyes corresponds to a defense.

Attackers are legitimate network agents that can \mathii{send} and \mathii{receive} messages, derive new messages by analyzing (e.g.\ decomposing)  known messages, obtain messages transiting
on the network (\mathii{spy}) and remove them so that they do not reach
their intended receiver (\mathii{erase}). Attackers can also partially
impersonate other agents, by \mathii{inject}ing messages under a false
identity; we represent impersonification with the notation $E(X)$, where $E$ is the impersonator and $X$ is the identifier of the impersonated agent.
This set of abilities
describes agents who have control over almost all facets of a
communication; their characteristic limitation is that they cannot
violate cryptography (we assume perfect cryptography). Note that further
rules could be added in Table~\ref{OurDYmodel} for other forms of encryption, digital signatures, hashing, creation of nonces and other fresh data, and so on.

The most significant feature concerns spying, represented through three
rules. For conceptual clarity, we explicitly pair an \mathii{erase}-rule
with the \mathii{injection}-rule, to emphasize that an attacker can
modify messages (by erasing them and injecting a substitute) or send
messages under a false identity (partial impersonification). Our
attackers can employ three different \mathii{spy} rules, adapted to formalize the
fact that attackers do not pay attention to all of the traffic on the
network. The \mathii{spy} rules rely on an interpretation for ``send''
that is modified with respect to the denotational semantics
in~\cite{Caleiro200688}, to reflect the attentional focus of attackers.
The default \mathii{spy} is the \rspy{}: only the messages involving
known agents in both sender and receiver roles, regardless of hypotheses
on their honesty, become part of the attacker's dataset. Note that in
our model what matters is the actual sender and not the declared sender
(\trueid). This mechanism prevents total impersonification and allows
filtering messages on the basis of the agent's attentional choices.

The attentional filter we use is meant as a choice of the agents and not
as a constraint to which they are subject; therefore, it must be
possible to expand the set of agents of interest. This role is fulfilled
by the two exploratory \mathii{spy} 
rules in Table~\ref{OurDYmodel},
\text{\ispy} and \text{\ospy}. Attackers have the option of accepting or
rejecting the newly discovered identifier $X$, on the basis of the
predicate \ofinterest{X}, which models the decisional process for
attention.

Note that an attacker cannot apply any of the \mathii{spy} rules to
obtain the message $m$ without knowing the identifier of at least one
between $m$'s sender and $m$'s intended receiver. By not providing a
``generalized spy'' rule to waive this requirement, we ensure that
$(D_{E}^{0} \: \cap$ \mathii{Agents} $= \emptyset)$ implies that for all
$i$, $(D_{E}^{i} \: \cap$ \mathii{Agents} $= \emptyset )$. Although $E$
can augment its knowledge base $D_E$ indefinitely --- through internal 
message generation and the synthesis rules  
\mathii{Comp} and \mathii{Encr} ---,
$E$'s network activity is in fact null. One such $E$ is a \emph{dummy
attacker}, whose usefulness becomes apparent when considering that proof
of reductions for non-collaboration can involve progressively migrating
identifiers from an attacker's dataset, until the attacker himself
reduces to the dummy attacker.

An attacker's dataset $D_E$ consists of (i) messages that have transited
through the network and that have been successfully received, analyzed
or spied and (ii) identifiers of the agents to whom the attacker is
attentive. The set \attentive{E} of identifiers of interest to $E$ is
further partitioned into three sets: the set \honest{E} of agents
believed\footnote{We do not attach any doxastic interpretation
to the beliefs we consider in this paper.} to be honest, the set
\dishonest{E} of agents believed to be attackers, and the set
\notknow{E} of agents whose attitude is unknown in $E$'s eyes. Note that
differently from $D_{net}$,
agent datasets do not contain triplets (\triple), but only
messages or identifiers.

Once a new identifier $X$ enters the knowledge base of attacker $E$, $E$
establishes a belief about the honesty of $X$ and places the identifier
in one of the sets \honest{E}, \dishonest{E} or \notknow{E}. 
We do not enter details on how the agents initially build their knowledge base and
establish their belief about the attitude of other known agents. In
fact, this classification is meant to be dynamic. Agents are on the
watch for suspicious messages, which may indicate that an attack is
ongoing or may reveal that a certain agent is dishonest. Dynamically
adapting their beliefs about the honesty of other agents allows the
agents to gather important information during single protocol runs. The
agents we wish to consider are \emph{smart}: they always employ the
available strategic information.

Attackers do not have automatic access to triplets that relate sender,
message and receiver. They must infer key pieces of information on the
basis of the identifiers of the agents to which they are attentive, and
attempt to relate the identifiers to the messages they spy. Inference is
easier if attackers use only the \rspy{} rule and keep the set of known
agents small. The difficulty of inference rises with the number of
attackers in the set \attentive{E}.

\subsection{Network model} \label{sec:evolution}

All the operations that can change the state of the network dataset $D_{net}$ 
(\mathii{send}, \mathii{receive}, \mathii{inject} and \mathii{erase})
are termed \emph{actions}, whereas we consider \mathii{spy} simply as an operation: although it requires interacting with the network, it does not change its state.
Messages in transit are inserted in the network dataset $D_{net}$, where
attackers can spy them before they are delivered to their intended
receivers. Contextually to delivery, the message is removed from the
dataset. Messages transit on the network dataset in the form of triplets
of the type \triple. As a consequence of message delivery or deletion, $D_{net}$ is non-monotonic by construction.

The sequence of actions that takes place during a protocol run is
enumerated and used to index the evolution of the network dataset
$D_{net}$; the index of $D_{net}^i$ is shared with all the proprietary datasets $D_{X}^{i}$, whose states are synchronized accordingly. 
$D^i_{net}$ is the state of the network dataset \emph{after the i-th action}.

Customarily, evolutions are indexed per transition (per rule application), rather than per action. Our chosen indexing strategy reflects three needs: (1) allowing agents to fully analyze newly acquired messages without having to keep track of the number of internal operations performed; (2) supporting a form of competition between attackers for access to the network; (3) supporting a form of concurrence. 

Ideally, all attackers act concurrently. However, the state transitions for the network must be well-defined at all times, even if attackers try to perform conflicting actions, such as spying and deleting the same message in transit. To impose a measure of order, we introduce a \emph{network handler}, whose task is to regulate the selection of the next action and implement the dependencies between selected action and knowledge available to each attacker; through the network handler, it is also possible to keep the system evolution in accordance with additional constraints, modeling for example information sharing within specific subsets of agents and network topology.

As soon as the state of the network changes (e.g.\ as a result of \mathii{inject} or \mathii{send}), the network handler passes the new triplet to each attacker, who then \emph{simulates} spying and decides on whether to request erasing the message or injecting a new one as a consequence, in accordance with his strategy. The network handler interprets the application of the spy-rules, the inject-rule and the erase-rule as requests and selects the next action from the set of requests. Message deletion, when requested by any attacker, is always successful.

The outcome of the process governed by the network handler is described through the function \cansee{}, which returns a subset of \mathii{Eves}, highlighting the identifiers of the attackers who can spy ``before'' the message is erased from $D_{net}$. The set of agents
described by \cansee{} contains at least the identifier of the
attacker whose erase request was served. 

If the network handler does not receive any erase-requests, all attentive attackers can acquire the message. If one or more erase-requests are present, the network handler erases the message and confirms success in spying only for a subset of attentive attackers. 
If an attacker is not in \cansee{}, the prior (simulated) spy is subject to rollback, along with all internal operations that have occurred since the last confirmed action. 
If no requests are received from attackers, the network handler oversees message delivery or selects actions requested by honest agents.

Although the formulation of \cansee{} in terms of access time is intuitive, the reason why we favor this mechanism is that time-dependent accessibility is not the only situation it can model. The function can be instantiated to model strategic decision-making and information-sharing, or to capture a particular network topology. In realistic attack scenarios, knowledge of a message that has been erased may depend more on cooperation and information-sharing than on timing. 
For example, if $E_j$ is sharing information with $E_k$ (but not viceversa), whenever $E_j$'s erase requests are served $E_k$ is automatically in \cansee{}.

The network handler is not an intelligent agent. Specifying its behavior and instantiating
the function \cansee{} corresponds to configuring the particular
network environment in which the agents are immersed
(i.e.~\cansee{} is a configurable parameter of our model).

As a result on the network handler and of our chosen indexing strategy, several internal operations can occur in a proprietary dataset between consecutive states, whereas only a single action separates consecutive states of the network dataset. Attackers determine the next
state of the network dataset with priority with respect to the actions of honest agents.

In Table~\ref{AliceAndBob}, we formalize within our model operations in the Alice\&Bob notation used in Section~\ref{sec:case-study}; we write $E_{I}(Y)$ to denote the subset
of \mathii{Eves} who spy message $m$ addressed to $Y$, at least one of which has requested $m$
to be erased.

With reference to Table~\ref{AliceAndBob}, note that the $(i+1)^{\text{th}}$ action is requested when the state of the network is $D^i_{net}$ and agent datasets are $D_X^i$; thus, the sender $X$ must already know in $D_X^i$ both the message $m$ and the identifier of the intended recipient $Y$. The message correctly transits on $D_{net}^{i+1}$, immediately after being sent. The (i+2)th action is either \mathii{receive} (first two cases) or \mathii{erase} (last case). the availability of $m$ to attackers is conclusively decided after the network handler selects the (i+2)th action, and thus pertains to $D_{W}^{i+2}$. 

\begin{table}[t!]
\begin{center}
\scalebox{0.65}{
\begin{tabular}{cl}
\hline
\puntello 
$\quad (i+1)^{\text{th}}$ action $\quad$ & Formalization\\
\hline 
\puntelloo
{\footnotesize $X \to Y \: : \: m$}
& {\footnotesize $m \in D_{X}^{i} \quad \text{and} \quad Y \in D_{X}^{i}$ } \\
\puntelloo
& {\footnotesize $<X,m,Y> \in D_{net}^{i+1} \quad \text{and} \quad <X,m,Y> \not\in D_{net}^{i+2}$ } \\
\puntelloo
& {\footnotesize $m \not\in D_W^{i+2},$ where $ W\not\in$~\cansee{<X,m,Y>, i+1} } \\
\puntelloo
& {\footnotesize $m \in D_{Y}^{i+2}$ } \\
\hline
\puntelloo
{\footnotesize $E(X) \to Y \: : \: m$}
& {\footnotesize $m \in D_{E}^{i} \quad \text{and} \quad X \in D_{E}^{i} \quad \text{and} \quad Y \in D_{E}^{i}$ } \\
\puntelloo
& {\footnotesize $<E(X),m,Y> \in D_{net}^{i+1} \quad \text{and} \quad <E(X),m,Y> \not\in D_{net}^{i+2}$ } \\
\puntelloo
& {\footnotesize $m \not\in D_W^{i+2},$ where $ W \not\in$ \cansee{<X,m,Y>, i+1} } \\
\puntelloo
& {\footnotesize $m \in D_{Y}^{i+2}$ } \\
\hline
\puntelloo
{\footnotesize $X \to E_{I}(Y) \: : \: m$}
& {\footnotesize $m \in D_{X}^{i} \quad \text{and} \quad Y \in D_{X}^{i}$ } \\
\puntelloo
& {\footnotesize $<X,m,Y> \in D_{net}^{i+1} \quad \text{and} \quad <X,m,Y> \not\in D_{net}^{i+2}$ } \\
\puntelloo
& {\footnotesize $m \in D_W^{i+2},$ where $ W \in I \: \text{and} \: I \subseteq$  \cansee{<X,m,Y>, i+1} } \\
 \hline \\
\end{tabular}
}
%
\caption{Representation of operations in Alice\&Bob notation.}
\label{AliceAndBob}
\end{center}
\end{table}

\subsection{Attacker goals and agent disposition}

The notion of cooperation between agents can be viewed from at least two
perspectives of interest: sharing of information and sharing of success.
The notion of attacker cooperation classically employed in protocol
analysis encompasses both aspects, as it states the first while assuming
that the second holds.

In this paper, we examine attackers that exhibit, with respect to
cooperation, the behavior we call \emph{complete non-collaboration}:
agents voluntarily abstain from sharing information and do not consider
their goals as met if they do not succeed in attacking.
The \emph{disposition} of attacker $E_1$ towards $E_2$ belongs to one of
the following basic classes: active collaboration, passive
collaboration, competition and conflict\footnote{In active and passive
collaboration there is a common goal to be pursued; the difference lies
in choosing a strategy that helps another vs.\ choosing a strategy that
does not hinder another. In conflict scenarios, the primary focus of
interest is the attackers, rather than the protocol.}. The focus of this
paper is on competition -- a situation in which the goal is successfully
attacking the protocol, regardless of the disposition of other agents.
From the perspective of a competitive attacker, other attackers are not
of interest per se: they are relevant factors because they are sources
of interference. If some interference is detected while carrying out an
attack, a competitive attacker will take countermeasures, attempting to
negate potentially adverse effects.

Sets of agents that are homogeneous with respect to disposition can be
used to define scenarios of interest. In the case study below, we explore a simple characteristic scenario composed of two
competitive attackers; we aim to bring into focus the mechanisms by
which two attackers can affect each other's success.

\section{A case study: the Boyd-Mathuria Example} 
\label{sec:case-study}

A dishonest agent, aware that other independent attackers may be active
on the network, will seek to devise suitable novel attacks, so as to
grant himself an edge on unsuspecting competitors. As the mechanics of
interaction and interference between attackers have not been
exhaustively studied in literature yet, it is not known a priori
how to systematically derive an attack behavior of this type. 

In the following case study, we start from a simple protocol for which a
vulnerability is known; we devise for the known (``classical'') attack a
variant that explicitly considers the possibility of ongoing independent
attacks. We describe a possible reasoning for a competitive attacker in
the context of the protocol's main features. Due to space limitations,
we give additional details about the case study in the appendix.

\begin{table}[t!]
\begin{center}
\scalebox{0.8}{
\begin{tabular}{cc}
\hline
{\bf BME} & {\bf Classical Attack}\\
\hline \\
\scriptsize
\!\!\!\!\!\!
$
\begin{array}{rll}
(1) & A \to S  & : A, B \\
(2) & S \to A  & :  \{\!| k_{AB} |\!\}_{k_{AS}}, \{\!| k_{AB} |\!\}_{k_{BS}} \!\!\!\!\!\! \\
(3) & A \to B  & : \{\!| k_{AB} |\!\}_{k_{BS}} \\
\end{array}
$
&
\scriptsize
$
\begin{array}{rll}
(1\phantom{'}) & A \to E(S)  & : A, B \\
(1') & E(A) \to S  & : A, E \\
(2\phantom{'}) & S \to A  & :  \{\!| k_{AE} |\!\}_{k_{AS}}, \{\!| k_{AE} |\!\}_{k_{ES}} \!\!\!\!\!\! \\
(3\phantom{'}) & A \to E(B)  & : \{\!| k_{AE} |\!\}_{k_{ES}} \\
\end{array}
$
\\\\
\hline
\end{tabular}
}
\end{center}
\caption{The Boyd-Mathuria Example protocol and a masquerading attack against it.}
\label{BME}
\end{table}

The protocol we consider as a case study is a key transport protocol
described as an example in~\cite{boydMathuria}; we name it as the
Boyd-Mathuria Example (BME), and present it in Table~\ref{BME} together
with a classical attack against it. BME relies on the existence of a
trusted third-party server $S$ to generate a session key $k_{AB}$ for
agents $A$ and $B$, where each agent $X$ is assumed to share a symmetric
secret key $k_{XS}$ with $S$.

$A$ is subject to a masquerading attack in which, at the end of a run of
BME, $A$ thinks that he shares a session key with the honest agent $B$,
while in fact he shares it with the attacker $E$. Subsequent
communication from $A$ addressed to $B$ is seen by $E$ through the
spy-rule and removed with an erase request: $E$ has successfully taken
$B$'s place.
This attack prevents $B$ from receiving \emph{any} communication from
$A$. Should the two agents have prior agreement that such a
communication was to take place, $B$ is in the position of detecting
that something has gone wrong. $E$ can prevent detection by staging a
dual man-in-the-middle attack.

If more than one attacker is active during a given protocol run,
simultaneous execution of the classical attack could lead to $A$
receiving multiple session keys as a response to his (single) request to
the server. This situation clearly indicates to $A$ that an attack is
ongoing. A competitive attacker $E_1$, wishing to prevent this situation
from occurring, could try removing from the network all the responses
from $S$ to $A$ that do not pertain to his own request. However, the
characteristics of the 
(non-redundant) cryptographic methods employed here do not allow
distinguishing $M_1 = \left(\{\!| k_{AE_1} |\!\}_{k_{AS}}, \{\!|
k_{AE_1} |\!\}_{k_{E_1S}}\right)$ (to let through) from $M_2 = \left(
\{\!| k_{AE_2} |\!\}_{k_{AS}}, \{\!| k_{AE_2} |\!\}_{k_{E_2S}}\right)$
(to block). $E_1$ can recognize the format of $M_1$ and $M_2$ and can
successfully decrypt $M_1$ to recover $k_{AE_1}$; by decrypting $M_2$
with the key $k_{E_1S}$, $E_1$ can still recover a value, but different
from the previous one. Not knowing $k_{AE_1}$ a priori, the attacker is
not able to distinguish which of $M_1$ and $M_2$ contains the answer to
his request for a key with $A$.

As a consequence, the attacker $E_1$ is not able to know which messages
to remove in order to ensure that $A$ accepts $k_{AE_1}$ as a session
key to communicate with $B$. Competitive attackers cannot rely on step
(2) to enforce their attacks at the expense of their competitors;
furthermore, the probability of erasing all competing messages (while
letting one's own pass) decreases with the number of active attackers.
In this situation, it becomes fundamental for a competitive attacker to
gain exclusive access to the first message and gain control over the
messages that reach $S$, as opposed to the messages coming from $S$
\footnote{
Of course, $E_1$ could guess which message(s) to erase, but 
he would have the added difficulty of having to decide whether to let 
the first message pass without knowing how many other messages will transit, 
if any at all, and how many session keys were requested by $A$ (as opposed to by his competitor(s)).
}.

After spying the initiator's opening message, a competitive attacker
$E_1$ will therefore attempt to mount the classical attack, while
keeping watch for other messages that may be interpreted as attack
traces. Any transiting message of the type $(A, E_m)$ for which $E_m
\in$ \dishonest{E_1} is interpreted as another active attack; $E_1$
counters by requesting that the message be erased. If $E_m$ is in
\honest{E_1}, the message may be understood either as a message from $A$
--- who would be initiating a parallel session of the protocol to obtain
a second session key --- or as an indication that $E_m$ has been
incorrectly labeled as honest. In the first case, $E_1$ will let the
message through, as he has chosen to target specifically the session key
for the communication between $A$ and $B$; in the second case, he will
protect his attack by erasing the message. If $E_m$ is in \notknow{E_1},
$E_1$ can choose to either play conservatively and hypothesize the
dishonesty of $E_m$ or let the message through and interpret $E_m$ as
the culprit in case the current attack fails.

BME is such that at most one attacker $E_d$ can successfully mislead $A$
into accepting the key $k_{AE_d}$ as a session key to communicate with
$B$. Therefore, a successful attack automatically entails exclusivity of
success. An attack is successful if it goes undetected by the initiator
$A$. Our honest agents are intelligent and they make use of all
information available to perform in-protocol detection of attacks. With
respect to BME, a clear indication for $A$ consists in receiving
multiple responses from $S$ after a single session key request; if $A$
receives multiple responses, he concludes that there has been a security
violation and thus does not employ 
any of the keys so received
in his later communications with $B$ 
-- choosing to try a fresh run of the protocol instead. 
From the attackers' perspective, an ongoing
attack can be detected by observing a message of the type $(A, X)$
transiting on the network; however, the attack trace is ambiguous to
spying attackers and has to be interpreted on the basis of current
beliefs concerning the honesty of $X$. A last feature of interest is
that BME is rather friendly for attacker labeling. Decisional processes
can rely on at least some conclusive information on the identity of the
agents involved, because identifiers transit in the clear; attackers
would have to infer them otherwise.

We examine the outcome of attacks carried out in a non-collaborative
environment in six cases, corresponding to different conditions of
knowledge and belief for $E_1$ and $E_2$. Cases and attack traces are
summarized in Table~\ref{tracesBME}. In order to completely specify
agent behavior, 
we posit the following:
\begin{enumerate}
\item If an attacker $E$ spies $(A, E_m)$ with $E_m \in $ \honest{E} or $E_m \in $ \notknow{E}, he will not request that the message be erased. In the latter case, if $E$'s attack fails, $E_m$ is immediately placed in \dishonest{E}.
\item Both $E_1$ and $E_2$ spy the opening message 
 and are interested in attacking the current protocol run; this allows us to leave aside the trivial cases in which only one attacker is active for a given protocol run.
\item Due to space constraints, we detail only the cases in which $\mathit{canSee}$ for step (3) yields $\{E_1,E_2\}$. Cases in which only one of the attackers can access $A$'s response can be found in  appendix \ref{extendedBME}.
\end{enumerate}

\noindent \textbf{Case 1: $E_{1}$ and $E_{2}$ know each other as honest.}\\
$E_{1}$ and $E_{2}$ know each other's identifiers (i.e.\ they are paying attention to each other: $E_{1} \in D_{E_{2}} \text{ and } E_{2} \in D_{E_{1}}$), but they are both mistaken in that they have labeled the other as honest ($E_{1} \in $ \honest{E_2} and $E_{2} \in$ \honest{E_1}).
$E_1$ and $E_2$ are unaware of active competitors and mount the classical attack in steps $(1_1)$ and $(1_2)$. When the attackers spy two requests to the server transiting on the network, they both believe that $A$ wishes to request keys with the honest agents B and $E_j$. 
\\
\textbf{(1.T1)}: \emph{$S$ sends two messages before $A$ can address a message to $B$.}
With the messages in steps $(2_1)$ and $(2_2)$, $A$ receives two keys instead of the single key requested. $A$ now knows that at least one attacker is active and abandons the protocol without sending a message to $B$. The attackers do not spy the message they were hoping for (timeout) and acquire the certainty that at least another active attacker is around. The attackers can employ ad-hoc strategies to search for the mislabeled or unknown attacker. If the attackers are careful to keep track of the messages $(A, X)$  pertaining to a given session, they can make informed guesses as to whom, amongst the known agents, they might have mislabeled. 
\\
\textbf{(1.T2)}:\emph{ $A$ receives a reply from $S$, answers $B$ and stops listening.}
$A$ receives the messages he expects and closes the current session before receiving the second response from $S$. $E_1$ is successful in his attack, whereas $E_2$ believes that he has succeeded when he has, in fact, decrypted the wrong key. None of the agents have an opportunity for detection.
\\
\textbf{(1.T3)}:\emph{ $A$ receives a reply from $S$, answers $B$ and keeps listening.}
$A$ replies with the message in step $(3)$, resulting in both $E_1$ and $E_2$ believing that they have succeeded. However, after receiving $(2_2)$, $A$ detects the attack and abstains from employing $k_{AE_1}$ in his future communications with $B$. Thus, even if for different reasons, both attackers in fact fail. Furthermore, they both continue to hold their mistaken belief that the other attacker is in fact honest. 

$ $\\
\noindent \textbf{Case 2: $E_{1}$ and $E_{2}$ know each other as attackers.}\\
$E_{1}$ and $E_{2}$ know each other's identifier ($E_{1} \in D_{E_{2}}$ and $E_{2} \in D_{E_{1}}$) and have correctly understood that the other is behaving as a dishonest agent 
($E_{1} \in $ \dishonest{E_2} and $E_{2} \in $ \dishonest{E_1}).
Each attacker is aware of the presence of a competitor, which they have correctly labeled. Each attacker is attempting to gain exclusive access to the initial communication towards $S$ and to ensure that only his request reaches $S$. $E_1$ and $E_2$ erase each other's request to $S$. Within our model, no attacker can be certain that his message has been received by its intended receiver; the attackers may wish to replay step $(1_1)$ and $(1_2)$ if a message of the type $\{\!| k_{AE_{j}} |\!\}_{k_{AS}}, \{\!| k_{AE_{j}} |\!\}_{k_{E_{j}S}}$ is not spied on the network within a reasonable time. This option is marked with $(\cdot)^{+}$ in Table~\ref{tracesBME}. However, the active presence of the competitor ensures that no message reaches $S$. $A$ notices that an anomalous situation is occurring, because his request to the server is not being served in a reasonable time. $A$ interprets the situation as a denial-of-service attack and abandons the protocol.

\begin{table}[t!]
\begin{center}
\scalebox{0.8}{
\begin{tabular}{cc}
\hline
\footnotesize {\bf  T1: cases 1, 3, 4, 6} & 
\footnotesize {\bf  T2 and [T3]: cases 1, 3, 4, 6}\\
\hline \\
\scriptsize
$
\begin{array}{rll}
(1\phantom{_{1}}) & A \to E_{1,2}(S)  & : A, B \\
\downarrow(1_{1}) & E_{1}(A) \to S  & : A, E_{1}  \\
\uparrow(1_{2}) & E_{2}(A) \to S  & : A, E_{2} \\
(2_{1}) & S \to A  & : M_{1} \\
(2_{2}) & S \to A  & :  M_{2} \\
\end{array}
$
&
\scriptsize
$
\begin{array}{rll}
(1\phantom{_{1}}) & A \to E_{1,2}(S)  & : A, B \\
\downarrow(1_{1}) & E_{1}(A) \to S  & : A, E_{1} \\
\uparrow(1_{2}) & E_{2}(A) \to S  & : A, E_{2} \\
(2_{1}) & S \to A  & :  M_{1} \\
(3\phantom{_{1}}) & A \to E_{1,2}(B)  & : \{\!| k_{AE_{1}} |\!\}_{k_{E_{1}S}} \\
\phantom{\{\!|} \left[ 
(2_{2}) \right. & S \to A  & :  M_{2} \left] 
\right. \\
\end{array}
$
\\ 
&\\
\hline
\footnotesize {\bf T4: case 2} & 
\footnotesize {\bf T5: case 5} \\
\hline \\
\scriptsize
$
\begin{array}{rll}
(1\phantom{_{1}})\phantom{^{+}} & A \to E_{1,2}(S)  & : A, B \\
\downarrow(1_{1})^{+} & E_{1}(A) \to E_{2}(S)  & : A, E_{1} \\
\uparrow(1_{2})^{+} & E_{2}(A) \to E_{1}(S)  & : A, E_{2} \\
\end{array}
$
&
\scriptsize
$
\begin{array}{rll}
(1\phantom{_{1}}) & A \to E_{1,2}(S)  & : A, B \\
\downarrow(1_{1}) & E_{1}(A) \to E_{2}(S)  & : A, E_{1} \\
\uparrow (1_{2}) & E_{2}(A) \to S  & : A, E_{2} \\
(2\phantom{_{1}}) & S \to A  & :  M_{2} \\
(3\phantom{_{1}}) & A \to E_{1,2}(B)  & : \{\!| k_{AE_{2}} |\!\}_{k_{E_{2}S}} \!\!\!\!\!\! \\
\end{array}
$
\\ 
&\\
\hline
\hline
\multicolumn{2}{c}{
\rule[-2 mm]{0mm}{0.7 cm}
Where: 	\scriptsize		
$M_{1} = \{\!| k_{AE_{1}} |\!\}_{k_{AS}}, \{\!| k_{AE_{1}} |\!\}_{k_{E_{1}S}} \:,\quad
  M_{2} = \{\!| k_{AE_{2}} |\!\}_{k_{AS}}, \{\!| k_{AE_{2}} |\!\}_{k_{E_{2}S}}$\!\!\!\!
}
\\
\hline
\end{tabular}
}
\end{center}
\caption{Traces for non-collaborative attacks against BME. Traces are exhaustive: $E_1$ and $E_2$ have priority over honest agents and $S$ is honest. Arrows: relative order between $(1_1)$ and $(1_2)$ is irrelevant in determining the outcome.}
\label{tracesBME}
\end{table}

$ $\\
\noindent \textbf{Case 3: $E_{1}$ and $E_{2}$ are unaware of each other.}\\
$E_{1}$ and $E_{2}$ are unaware of the other's presence -- i.e.\ they are not paying attention to the other's activity ($E_{1} \notin D_{E_{2}}$ and $E_{2} \notin D_{E_{1}}$). 
Subcases follow closely those described for case $1$ above. The only
significant difference concerns detection for trace T1: here the
attackers must employ exploratory strategies (\ispy{} or \ospy{}),
because they failed to spy an additional message of type $(A, E_m)$
transiting on the network. The failure to observe such a message is a
strong indicator that the competitor's identifier is unknown. In
2-attacker scenarios this is the only legitimate conclusion, whereas
with three or more attackers this situation may also arise from the
interplay between erase and spy operations.

$ $\\
\noindent \textbf{Case 4: $E_{2}$ knows $E_{1}$ as honest.}\\
Only one out of the two attackers $E_{1}$ and $E_{2}$ is paying attention to the other and knows his identifier. Here we consider $E_{1} \in $ \honest{E_2} and $E_{2} \notin D_{E_{1}}.$
Regardless of the order in which steps $(1_1)$ and $(1_2)$ occur, the attacker in disadvantage $E_1$ does not spy the message at step $(1_2)$; $E_2$ does spy $(1_1)$ but, trusting his judgement on $E_1$'s honesty, does not request it to be erased. As a consequence, similarly to case 1, the traces follow schemes T1, T2 and T3. Significant differences concern detection in T1: $E_1$ detects the presence of an unknown attacker, whereas $E_2$ learns of a mislabeled or unknown attacker. The successful attackers in traces T2 and T3 are those whose requests to $S$ are served first; knowledge does not affect the outcome.

$ $\\
\noindent \textbf{Case 5:  $E_{2}$ knows $E_{1}$ as dishonest.}\\
Only one out of the two attackers $E_{1}$ and $E_{2}$ is paying attention to the other and knows his identifier. Here we consider $E_{1} \in $ \dishonest{E_2} and $E_{2} \notin D_{E_{1}}$
Regardless of the order in which steps $(1_1)$ and $(1_2)$ occur, $E_1$ does not spy the message at step $(1_2)$ and $E_2$ uses a direct attack against the competitor. $E_2$ removes $E_1$'s request to the server and remains the only attacker in play, leading $A$ into accepting $k_{AE_2}$ as a session key. $E_1$ does not have an opportunity to detect the competitor. 

$ $\\
\noindent \textbf{Case 6: $E_{2}$ knows $E_{1}$, but he is unsure of $E_{1}$'s honesty.}\\
Only one out of the two attackers $E_{1}$ and $E_{2}$ is paying attention to the other and knows his identifier. Here we consider $E_{1} \in $ \notknow{E_2} and $E_{2} \notin D_{E_{1}}$. This case reduces to case 4, with the only difference that $E_2$ is testing the dishonesty of $E_1$, instead of believing his honesty. Whenever $E_2$ realizes that he has failed his attack, 
he adds $E_1$ into \dishonest{E_2} and deletes it from \notknow{E_2}. 

$ $\\
\noindent \textbf{General considerations.}\\
In traces T2 and T3, the winning attacker is the one whose request is served first by $S$. $S$ is an honest agent but it is not constrained to answering requests in the exact order in which they are received. Attackers do not have control over which requests are served first, although this factor determines whether they cannot do better than acquire the wrong key. Attackers realize in-protocol that they have failed only when they cannot spy a response from $A$, i.e.\ when they do not acquire any keys. Post-protocol detection, on the other hand, can occur also when an attacker with a wrong key attempts to decrypt the later communications addressed by $A$ to $B$. 

The case study highlights that, if $A$ keeps the session open for a
reasonable time after step (3), he can improve his chances of
discovering that the key is compromised. This is a simple strategy that
is beneficial and does not depend on the particular protocol.
Furthermore, when $A$ receives two answers from $S$ in response to his
single request, he now has two keys -- at least one of which is shared
with an attacker. If honest agents are immersed in a retaliatory
framework \cite{Bella03,retaliation}, such keys can be used to identify
attackers, to feed them false information or, in general, to launch
well-aimed retaliatory attacks.

\section{Defending vulnerable 
protocols against attacks}
\label{sec:defense}

Key exchange protocols are amongst the most used cryptographic
protocols. It is a common security practice to establish a secure
channel by first exchanging a session key and then using it to
authenticate and encrypt the data with symmetric cryptography. The
security of all communications occurring during a session rests on the
integrity of the key. In this context, it is not important per se that a
key has been acquired by an attacker: what matters is whether a
compromised key is used. Rather then on preventing the acquisition of a
session key from ever occurring, the focus is on detecting that the key
has been compromised -- so as to prevent an attack from spreading to the
entire session traffic.

If a protocol is vulnerable, a single DY attacker will succeed with
certainty. However, if attacks to the same protocol are carried out in a
more complex network environment, success is not guaranteed. As shown in
the case study, in competitive scenarios with equal-opportunity
attackers it is not possible for a given attacker to ensure that an
attack is successful under all circumstances. The outcome depends on the
strategy and knowledge conditions of all the active agents, on the
visibility of erased messages to other attackers ($\mathit{canSee} \ne
\{E_1, E_2\}$) and on the order with which $S$ processes requests. In a
sense, the presence of an independent active attacker constrains the
success of otherwise sure-fire attacks.

This principle can be exploited to facilitate detection of attacks
against vulnerable 
protocols. Honest agents should not, in principle, be
informed of the specific attack trace to which they are vulnerable.
Hence, if honest agents can perform detection at all, it has to be on
the basis of flags that are independent of the specific attack trace --
and, in general, independent also of the protocol in use. Such flags
encode \emph{local} defense criteria and can be as simple as realizing
that no answer has arrived within a time considered reasonable or
realizing that two (different) answers have been sent in response to a
single request.

The basic idea is constructing a network agent that causes
protocol-indepen\-dent flags to be raised -- via 
deliberate 
interference with ongoing attacks. In addition, one such \emph{guardian agent} is
formally an attacker, and can therefore be configured with knowledge of
the attack trace(s). The guardian's task can be formulated as raising
protocol-independent flags in correspondence to protocol-dependent
indicators.

\begin{table}[t]
\begin{center}
\begin{tabular}{c|cccc}
\hline
\footnotesize \!\!\!\!\!
\textbf{~~canSee~~}	 \!\!\!\!\! & 
\footnotesize \!\!\!\!\!
\textbf{~~Cases~~}  \!\!\!\!\! & 
\footnotesize \!\!\!\!\!
\textbf{~~Case 2~~}  \!\!\!\!\!& 
\footnotesize \!\!\!\!\!
\textbf{~~Case 5}: ~  \!\!\!\!\!& 
\footnotesize \!\!\!\!\!
\textbf{~Case 5}: ~  \!\!\!\!\!\\
\footnotesize
	& 
\footnotesize
\textbf{~~1,3,4,6~~} & 
\footnotesize
 & 
\footnotesize
\textbf{~~$E\in$\dishonest{G}~} & 
\footnotesize
\textbf{~ $G\in$\dishonest{E}~} \\
\hline
\puntello
\footnotesize $\{ E, G\}$ & \canHelp$^{+}$ & \helpAll & \helpAll & \\
\puntello
\footnotesize $\{ G\}$ & \helpAll & \helpAll & \helpAll & \helpAll \\
\puntello
\footnotesize $\{ E\}$ & \canHelp$^{+}$ & \helpAll & \helpAll & \\
\hline
\end{tabular}
\end{center}
\caption{
Effects of introducing a guardian $G$ for BME when attacker $E$ is active. $G$ operates according to the same strategy as the attackers in the case study. $G$'s active interference results in $A$ detecting attacks always (\helpAll), sometimes (\canHelp), always if $A$ commits to listening after step (3) ($^{+}$). The guardian is progressively more effective the more his beliefs and knowledge reflect the actual set of attackers. $G$ can be effective even when he is not aware of $E$'s presence.
} 
\label{table:GuardianBME}
\end{table}

By using such an ad-hoc competitor as defense, it is possible, in some
cases, to allow detection of otherwise-undetectable attacks. If no flag
is raised for $A$, the guardian may be the only attacker at work. In
this case, no ill-intentioned attacker has successfully concluded an
attack; from the standpoint of $A$, actual security is not affected. A
guardian is a practical solution even when it is not all-powerful: any
attack detected by $A$ thanks to the guardian's active presence is an
improvement in security. In Table~\ref{table:GuardianBME}, we show the
effects of introducing a guardian $G$ for BME, configured as the
attackers in the case study. 
It is not necessary to demand that the 
guardian monitor all traffic -- which is unrealistic at best; on the 
other hand, all monitored traffic enjoys partial protection. 

Attacks failing are, by themselves, markers that there are other
dishonest agents at work; this fact can be used by the guardian $G$ as a
basis for further detection, possibly on behalf of honest agents. Then
guess-and-test strategies can be used to acquire an understanding of the
second attacker's identity; a rudimentary example is the strategy used
by our attackers for BME when they spy $(A, E_m)$ and $E_m \in$
\notknow{}. Across multiple iterations of the attack procedure and
under different hypotheses concerning (\honest{G}, \dishonest{G},
\notknow{G}), the attacker's identity will eventually be revealed.

In actual scenarios, protocols are implemented through programs in the users' computers. It is very difficult to force users to stop using a protocol as soon as a vulnerability is discovered. The more widespread the protocol, the more difficult it is to ensure that it quickly goes out of use. Two aspects are important: that every user (a) is informed of the new vulnerability and (b) takes action in switching to a secure protocol. Statistics on software upgrades are an unfortunate example of this type of issue.

By designing the user-end software to inform the user of a security failure whenever protocol-independent flags are raised, a guardian can help solve the notification issue as well as raise the likelihood that the user will take action and upgrade. When the weakness in
the protocol is understood, it may be a cost-effective investment to design a guardian with an effective interference strategy, so as to facilitate restoring network security.

\section{Conclusions and future work}
\label{sec:conclusions}

The traditional goal of protocol analysis is discovering attacks, to
prompt replacing a vulnerable protocol with an improved and more secure
one. Reductions are centered on attacks, either to reduce the search
space for attacks (e.g.~\cite{DTL,MillenDenker02,MVB-JCS10}) or to
reduce the number of agents (e.g.~\cite{DTL,comon03twoAgentsEnough}). In
particular, if there exists an attack involving $n$ collaborating attackers, then
there exists an ``equivalent'' attack involving only one. Within this
perspective, it is known that $n$-DY attackers equal in attack power a
single DY attacker, and that the same can be said of Machiavelli-type
attackers~\cite{MR712376,Syverson00dolev-yaois}. As a result, an
exhaustive search for attacks can be performed in a reduced-complexity
model.

On the other hand, within our proposed approach the goal of analysis is finding a strategy to defend the system against existing attacks, rather than identifying vulnerabilities to prompt redesigning the protocol. We would be performing protocol analysis to identify possible \emph{defenses}, rather than attacks.

In the case study, we have shown a counterexample to the statement: ``if
there exists a defense against an attack in a 2-attacker scenario, then
there exists an equivalent defense in a 1-attacker scenario''. This
statement mirrors the classical result on $n$-to-$1$ reducibility and
the counterexample shows that exhaustive searches for defenses against
attacks cannot be carried out in reduced-complexity settings, as 
they require at least two attackers. 

Having chosen vulnerable protocols, in a single-attacker situation there
is no pro\-to\-col-independent indicator that could be used by honest agents
to become aware that security has been compromised. If there is a single
attacker, no simple defense is possible and the protocol inevitably
fails its security goals. On the other hand, by exploiting an ad-hoc
competitor (the guardian) as a defense, in certain conditions we can
successfully raise protocol-independent indicators of ongoing attacks
and protect the system. Introducing an appropriate guardian procedure as
soon as new attacks are discovered can mitigate the consequences of flawed 
protocols still being in use.

Along the line of work presented in this paper, we have investigated two
additional simple protocols as case studies: the Shamir-Rivest-Adleman
three-pass protocol, which differs significantly from BME in that
success is not necessarily exclusive, and the Beller-Yacobi protocol,
which requires interacting with a second honest agent to carry out a
masquerading attack. The goal of these investigations
 is to bring into focus how the salient
features of each protocol are reflected in the possible mechanisms of
interference. 
The first case study is available as additional material 
in appendix \ref{casestudySRA3P}.
A second topic of interest is evaluating (i) whether the
mechanisms of interaction highlighted in two-attacker scenarios are
directly portable to situations with more than two non-collaborating
attackers, (ii) whether they require ad-hoc generalization and (iii) whether new
types of interaction emerge when more than two dishonest agents are
active. We are investigating this in more detail, along with a
(semi-)automatic implementation of our approach.

\subsection*{Acknowledgements}

The work presented in this paper was partially supported by the
FP7-ICT-2007-1 Pro\-ject no. 216471, ”AVANTSSAR: Automated Validation of
Trust and Security of Service-oriented Architectures” (www.avantssar.eu)
and by the FP7-ICT-2009-5 Project no. \\257876, ``SPaCIoS: Secure
Provision and Consumption in the Internet of Services''
(www.spacios.eu). We thank Davide Guardini for his constructive
comments.

\appendix 

\section{Extended tables for BME}\label{extendedBME}

In this appendix, we present a detailed view of the outcome of an attack
carried out against BME and involving only the non-collaborative
attackers $E_1$ and $E_2$. Refer to Section~\ref{sec:case-study} for a
definition of BME, attacker behavior against BME, attack traces and
cases.

Note that in cases 1, 2 and 3 (shown in Table~\ref{table:BME_extended}), $E_j$'s request is the $j$-th served by $S$. 
In cases 4, 5 and 6, $E_2$ is the attacker with knowledge advantage. For clarity, for cases 4 and 6 (see Table~\ref{table:BME_extended2}) we mark as $E_j$* the case in which $E_j$'s request is served \emph{first} by $S$. In case 5, $E_2$'s request is the only served and the distinction is unnecessary. 

A competitive attacker $E$ attacking BME can:
\begin{itemize}
 \item succeed and compromise a key that $A$ will use;
 \item fail and realize it (by timeout);
 \item fail without realizing it, by acquiring the wrong key;
 \item fail without realizing it, even though $E$ acquired the right key.
\end{itemize}
Honest agents under attack can:
\begin{itemize}
 \item detect the attack and abandon the protocol before carrying out step (3);
 \item realize that the key has been compromised and keep safe by not using it;
 \item fail to detect an attack but use their keys safely, because all attackers have failed to acquire the correct key;
 \item use a compromised key.
\end{itemize}
Attackers who realize their failure can infer the following:
\begin{itemize}
 \item[$\alpha$] \emph{Mislabeled or unknown attacker.} The attacker spies two messages from $S$ and none from $A$ in response; he deduces that $A$ had opened a single session and that at least one request to $S$ (in addition to his own) was an attack. The attacker realizes that he has either mislabeled as honest one of the active attackers or that an unknown competitor is active. 
 \item[$\beta$] \emph{Unknown attacker.} The attacker spies two messages from $S$ and none from $A$ in response; he deduces that $A$ had opened a single session and that at least one request to $S$ (in addition to his own) was an attack. However, he has seen no additional requests of the type $(A, X)$ transit on the network; the attacker realizes that an unknown competitor is active on the network.
 \item[$\gamma$] \emph{Missed message: mislabeled or unknown attacker.} The attacker spies only one message from $S$ but no reply from $A$; all messages from $S$ that successfully reach $A$ are seen, so the attacker deduces that he has missed $A$'s response. Thus, an active competitor (mislabeled or unknown) has erased it, preventing the attacker from acquiring it through the spy rule. 
\item[$\delta$] \emph{Missed message.} Similar to case $\gamma$. The attacker does not draw further conclusions because he is already aware of an active attacker that may have erased the message. 
\item[$\epsilon$] \emph{Suspect condemned.} The attacker $E$ has put to test the dishonesty of an agent X in \notknow{E} (the suspect). Failing the attack is interpreted as a confirmation that the suspect is dishonest: $X$ is placed into \dishonest{E}.
\end{itemize}

\begin{table}[t]
\newcommand{\mc}[3]{\multicolumn{#1}{#2}{#3}}
\begin{center}
\scalebox{0.55}{
\begin{tabular}{|cc|cccc|cc|c|}
\hline
\textbf{Trace} & \textbf{canSee} & \textbf{Agent} & \textbf{Result} & \textbf{Belief} & \textbf{Key} & \mc{2}{c|}{\textbf{Detection} } & \textbf{Guardian} \\
\hline
\textbf{T1} &  & × & × & × & × & \textbf{Case 1} & \textbf{Case 3} & × \\
\hline
 & -- & $E_1$ & failure & failure & none & $\alpha$ & $\beta$ & of help \\
× & × & $E_2$ & failure & failure & none & $\alpha$ & $\beta$ & of help \\
× & × & $A$ & safe & attack & not used & 2 keys & 2 keys & × \\
\hline
\textbf{T2} & \textbf{step (3)} & × & × & × & × & \textbf{Case 1} & \textbf{Case 3} & × \\
\hline
× & $\{E_1,E_2\}$ & $E_1$ & success & success & right & none & none & of help \\
× & × & $E_2$ & failure & success & wrong & none & none & no effect \\
× & × & $A$ & attacked & safe & broken & none & none & × \\
\hline
× & $\{E_1\}$ & $E_1$ & success & success & right & none & none & of help \\
× & × & $E_2$ & failure & failure & none & $\gamma$ & $\gamma$ & no effect \\
× & × & $A$ & attacked & safe & broken & none & none & × \\
\hline
× & $\{E_2\}$ & $E_1$ & failure & failure & none & $\gamma$ & $\gamma$ & of help \\
× & × & $E_2$ & failure & success & wrong & none & none & of help \\
× & × & $A$ & safe & safe & used & none & none & × \\
\hline
\textbf{T3} & \textbf{step (3)} & × & × & × & × & \textbf{Case 1} & \textbf{Case 3} & ×\\
\hline
× & $\{E_1,E_2\}$ & $E_1$ & failure & success & right & none & none & of help \\
× & × & $E_2$ & failure & success & wrong & none & none & of help\\
× & × & $A$ & safe & attack & not used & 2 keys & 2 keys & × \\
\hline
× & $\{E_1\}$ & $E_1$ & failure & success & right & none & none & of help \\
× & × & $E_2$ & failure & failure & none & $\gamma$ & $\gamma$ & of help \\
× & × & $A$ & safe & attack & not used & 2 keys & 2 keys & × \\
\hline
× & $\{E_2\}$ & $E_1$ & failure & failure & none & $\gamma$ & $\gamma$ & of help \\
× & × & $E_2$ & failure & success & wrong & none & none & of help \\
× & × & $A$ & safe & attack & not used & 2 keys & 2 keys & × \\
\hline
\textbf{T4} & -- & × & × & × & × & \mc{2}{c|}{\textbf{Case 2}} & × \\
\hline
~ & ~ & $E_1$ & failure & failure & none & \mc{2}{c|}{correct understanding} & of help \\
~ & ~ & $E_2$ & failure & failure & none & \mc{2}{c|}{correct understanding} & of help \\
~ & ~ & $A$  & safe & attack & none & \mc{2}{c|}{no answer: DoS} & × \\
\hline
\textbf{T5} & \textbf{step (3)} & × & × & × & × & \mc{2}{c|}{\textbf{Case 5}} & × \\
\hline
× & $\{E_1,E_2\}$ & $E_1$ & failure & success & wrong & \mc{2}{c|}{none} & no effect \\
× & × & $E_2$ & success & success & right & \mc{2}{c|}{correct understanding} & of help \\
× & × & $A$ & attacked & safe & broken & \mc{2}{c|}{none} & × \\
\hline
× & $\{E_1\}$ & $E_1$ & failure & success & wrong & \mc{2}{c|}{none} & of help \\
× & × & $E_2$ & failure & failure & none & \mc{2}{c|}{$\delta$} & of help \\
× & × & $A$ & safe & safe & in use & \mc{2}{c|}{none} & × \\
\hline
× & $\{E_2\}$ & $E_1$ & failure & failure & none & \mc{2}{c|}{$\gamma$} & no effect \\
× & × & $E_2$ & success & success & right & \mc{2}{c|}{correct understanding} & of help \\
× & × & $A$ & attacked & safe & broken & \mc{2}{c|}{none} & × \\
\hline
\end{tabular}
}
\end{center}
\caption{Outcomes of a competitive attack against BME involving the attackers $E_1$ and $E_2$ and the honest initiator $A$ \textbf{(cases 1, 2, 3 and 5)}. Traces are described in Table~\ref{tracesBME}; \cansee{} describes the set of attackers who spy the message sent by $A$ at step $(3)$; for each role, we report the actual result of the attack (result), if the agent believes he has succeeded or failed (belief) and whether he has acquired the right key, the wrong key or no key at all (key). When attackers realize their failure, they can infer the reason for failing as shown in the column Detection; the honest agent $A$ can detect ongoing attacks by receiving two answers from $S$ or none. In the last column, we show the result of introducing a guardian agent, playing the role in the corresponding row against an attacker playing the other role.}
\label{table:BME_extended}
\end{table}

\begin{table}[t]
\newcommand{\mc}[3]{\multicolumn{#1}{#2}{#3}}
\begin{center}
\scalebox{0.55}{
\begin{tabular}{|cc|cccc|cc|c|}
\hline
\textbf{Trace} & \textbf{canSee} & \textbf{Agent} & \textbf{Result} & \textbf{Belief} & \textbf{Key} & \mc{2}{c|}{\textbf{Detection}} & \textbf{Guardian} \\
\hline
\textbf{T1} & -- & × & × & × & × & \textbf{Case 4} & \textbf{Case 6} & × \\
\hline
× & × & $E_1$* & failure & failure & none & $\beta$ & $\beta$ & of help \\
× & × & $E_1$\phantom{*} & failure & failure & none & $\beta$ & $\beta$ & of help \\
× & × & $E_2$* & failure & failure & none & $\alpha$ & $\epsilon$ & of help \\
× & × & $E_2$\phantom{*} & failure & failure & none & $\alpha$ & $\epsilon$ & of help \\
× & × & $A$\phantom{*} & safe & attack & not used & 2 keys & 2 keys & × \\
\hline
\textbf{T2} & \textbf{step (3)} & × & × & × & × & \textbf{Case 4} & \textbf{Case 6} & × \\
\hline
× & $\{E_1,E_2\}$ & $E_1$* & success & success & right & none & none & of help \\
× & × & $E_1$\phantom{*} & failure & success & wrong & none & none & no effect \\
× & × & $E_2$* & success & success & right & none & none & of help\\
× & × & $E_2$\phantom{*} & failure & success & wrong & none & none & no effect \\
× & × & $A$\phantom{*} & attacked & safe & broken & none & none & × \\
\hline
× & $\{E_1\}$ & $E_1$* & success & success & right & none & none & of help \\
× & × & $E_1$\phantom{*} & failure & success & wrong & none & none & of help \\
× & × & $E_2$* & failure & failure & none & $\gamma$ & $\epsilon$  & of help \\
× & × & $E_2$\phantom{*} & failure & failure & none & $\gamma$ & $\epsilon$  & no effect \\
× & × & $A$\phantom{*} & attacked & safe & broken & none & none & ×  \\
\hline
× & $\{E_2\}$ & $E_1$* & failure & failure & none & $\gamma$ & $\gamma$ & of help \\
×  & × & $E_1$\phantom{*} & failure & failure & none & $\gamma$ & $\gamma$ & no effect \\
× & × & $E_2$* & success & success & right & none & none & of help \\
× & × & $E_2$\phantom{*} & failure & success & wrong & none & none & of help \\
× & × & $A$\phantom{*} & safe & safe & used & none & none & × \\
\hline
\textbf{T3} & \textbf{step (3)} & × & × & × & × & \textbf{Case 4} & \textbf{Case 6} & × \\
\hline
× & $\{E_1,E_2\}$ & $E_1$* & failure & success & right & none & none & of help \\
× & × & $E_1$\phantom{*} & failure & success & wrong & none & none & of help \\
× & × & $E_2$* & failure & success & right & none & none & of help \\
× & × & $E_2$\phantom{*} & failure & success & wrong & none & none & of help \\
× & × & $A$\phantom{*} & safe & attack & not used & 2 keys & 2 keys & × \\
\hline
× & $\{E_1\}$ & $E_1$* & failure & success & right & none & none & of help \\
× & × & $E_1$\phantom{*} & failure & success & wrong & none & none & of help \\
× & × & $E_2$* & failure & failure & none & $\gamma$ & $\epsilon$ & of help \\
× & × & $E_2$\phantom{*} & failure & failure & none & $\gamma$ & $\epsilon$ & of help \\
× & × & $A$\phantom{*} & safe & attack & not used & 2 keys & 2 keys & × \\
\hline
× & $\{E_2\}$ & $E_1$* & failure & failure & none & $\gamma$ & $\gamma$ & of help \\
× & × & $E_1$\phantom{*} & failure & failure & none & $\gamma$ & $\gamma$ & of help \\
× & × & $E_2$* & failure & success & right & none & none & of help \\
× & × & $E_2$\phantom{*} & failure & success & wrong & none & none & of help \\
× & × & $A$\phantom{*} & safe & attack & not used & 2 keys & 2 keys & × \\
\hline
\end{tabular}
}
\end{center}
 \caption{Outcomes of a competitive attack against BME involving the attackers $E_1$ and $E_2$ and the honest initiator $A$ \textbf{(cases 4 and 6)}. $E_j$*: $E_j$'s request at step $(1_i)$ is served by $S$ first. Traces are described in Table~\ref{tracesBME}; \cansee{} 
describes the set of attackers who spy the message sent by $A$ at step $(3)$; for each role, we report the actual result of the attack (result), if the agent believes he has succeeded or failed (belief) and whether he has acquired the right key, the wrong key or no key at all (key). When attackers realize their failure, they can infer the reason for failing as shown in the column Detection; the honest agent $A$ can detect ongoing attacks by receiving two answers from $S$ or none. In the last column, we show the result of introducing a guardian agent playing the role in the corresponding row against an attacker playing the other role.}
\label{table:BME_extended2}
\end{table}

\section{A case study: the Shamir-Rivest-Adleman Three-Pass Protocol} \label{casestudySRA3P}

The Shamir-Rivest-Adleman Three-Pass protocol (SRA3P), described in~\cite{Clark97asurvey}, has been proposed to transmit data securely on insecure channels, bypassing the difficulties connected to the absence of prior agreements between the agents $A$ and $B$ to establish a shared key. The security property targeted by SRA3P is confidentiality; if the message transmitted is interpreted as a session key, then the protocol can be considered as a \emph{key transport} protocol. 

SRA3P relies on the assumption that the kind of cryptography employed is commutative, i.e.\ that 
$\{\!| \{\!| M |\!\}_{K_{A}}|\!\}_{K_{B}}=\{\!| \{\!| M |\!\}_{K_{B}}|\!\}_{K_{A}}$ holds. We use the standard notation for symmetric cryptography to emphasize commutativity. The protocol consists in three message exchanges, as shown in Table~\ref{SRA3Pattacks}A.

The classical attack to SRA3P exploits $A$ as an oracle for the content of the message (Table~\ref{SRA3Pattacks}B). The attacker $E$ replaces the intended recipient $B$ in receiving the message and pretends to perform step (2) -- in actuality sending back the message $\{\!| M |\!\}_{K_{A}}$ without further encryption. 
$A$ continues according to the protocol and removes his key from the message, thus sending back the \emph{secret} $M$ without any encryption. We represent the message as $M^{*}$ to emphasize that $M$ transits in the clear without $A$ meaning it.

The classical attack is successful; however, it prevents the intended recipient $B$ from receiving any messages at all. In case the honest agents had prior agreement that an exchange was to take place, $B$ can detect that something has gone wrong. The classical attack is very strong against detection even in this case: after discovering $M$, the attacker $E$ impersonates $A$ and performs the protocol with $B$, de facto carrying out a complete man-in-the-middle attack. In this manner, the attack on SRA3P goes completely undetected and the attacker gains access to the secret key $M$.

Provided that some attacker answered $A$ in step (2) by sending $\{\!| M |\!\}_{K_{A}}$, it is sufficient to spy the message in step (3) to acquire the secret. In our set-up, any attacker attempting to erase a message is always successful in preventing honest agents from receiving it, but he is not necessarily successful in hiding it from other attackers (all attackers in $\text{canSee}(<A,M^{*},B>,i)$ have access to $M$).
In this situation, $E_2$ can prevent his competitors from acquiring the secret only by weakening their ability to identify the message $M^{*}$ as the true response of $A$ in step (3). A competitive attacker will therefore attempt to mislead his competitors by sending on the network fake messages that are in no way related to the information coming from the initiator $A$.


\begin{table}[ht] 
\begin{center}
\scalebox{0.7}{
\begin{tabular}{c|c}
\hline

{\bf (A)~~SRA3P } & {\bf (B)~~Classical Attack} \\
\hline 
&\\
$
\begin{array}{rll}
(1) & A \to B  & : \{\!| M |\!\}_{K_{A}} \\
(2) & B \to A  & : \{\!| \{\!| M |\!\}_{K_{A}}|\!\}_{K_{B}} \\
(3) & A \to B  & : \{\!| M |\!\}_{K_{B}} \\
\end{array}
$
&
$
\begin{array}{rll}
(1) & A \to E(B)  & : \{\!| M |\!\}_{K_{A}} \\
(2) & E(B) \to A  & : \{\!| M |\!\}_{K_{A}} \\
(3) & A \to E(B)  & :  M^{*} = M \\
\end{array}
$
\\
& \\
\hline
{\bf (C) Strong attack} & {\bf (D) Competitive attack}\\
\hline
&\\
$
\begin{array}{rll}
(1\phantom{'}) & A \to E_{1,2}(B)  & : \{\!| M |\!\}_{K_{A}} \\
(2\phantom{'}) & E_{1}(B) \to A  & : \{\!| M |\!\}_{K_{A}} \\
(3') & E_{2}(A) \to E_{1}  & :  M_{fake} \\
(3\phantom{'}) & A \to E_{1,2}(B)  & :  M^{*} \\
\end{array}
$
&
$
\begin{array}{rll}
(1 \phantom{'}) & A \to E_{1,2}(B)  & : \{\!| M |\!\}_{K_{A}} \\
(2\phantom{'}) & E_{1}(B) \to A  & : \{\!| M |\!\}_{K_{A}} \\
(3') & E_{2}(A) \to E_{1}(B)  & :  M_{fake} \\
(3\phantom{'}) & A \to E_{1,2}(B)  & :  M ^{*}\\
\end{array}
$
\\
& \\
\hline
\end{tabular}
}
\end{center} 
\caption[Attacks against SRA3P]{Attacks against the Shamir-Rivest-Adelman Three-Pass Protocol (SRA3P). $K_A$ and $K_B$ are private keys and cryptography is commutative. (A):~Protocol followed by honest agents. (B):~ Classical attack on SRA3P, employed by attackers when unaware of active competitors. (C):~Strong non-collaborative attack, employed by attackers when the competitor's identifier is known ($E_{2}$ knows that his competitor is $E_{1}$). (D):~Competitive attack, employed by attackers when aware of the existence of an active competitor but unsure of the competitor's identity ($E_{2}$ knows that he has a competitor but does not know that it is $E_{1}$).}
\label{SRA3Pattacks}
\end{table}

If the recipient of a fake message is expecting to receive $M^{*}$, he may be led into thinking that he has successfully carried out his attack. He may then stop spying the current run of the conversation between $A$ and $B$ and conclude that he has succeeded when in fact he has acquired the wrong ``secret'' $M_{fake}$. If, instead, the competitor $E_{1}$ is not following the classical attack and chooses to keep listening in on the conversation, he receives more than one message playing the role of $M^{*}$ and does not know which one has been sent by the honest agent $A$. 

The competitor faces a degree of uncertainty in identifying $M^{*}$ that is not present in the classical attack. The degree of uncertainty to which $E_{1}$ is subject can be increased arbitrarily by $E_{2}$, who can send multiple and unrelated fake messages, both before and after $M^{*}$ transits on the network. This style of attack grows in effectiveness as $E_{2}$ is better able to construct misleading fake messages. 

The success of this non-collaborative behavior in securing sole ownership of the secret depends critically on the listening behavior of the competitor: if $E_{1}$ stops spying network traffic as soon as a response is received, then it is critical for $E_{2}$ to send a fake message \emph{before} $A$'s reply; in case of success, the competitor fails to acquire the secret. If the competitor is actively listening past the reception of the first response, then $M^{*}$ is eventually acquired -- but not by itself: a situation of uncertainty arises. 

In classical settings, uncertainty does little more than affect the probability that an attack will be successful; however, if honest agents are immersed in a retaliatory framework, guessing the wrong $M^{*}$ and using it as a session key to communicate with $A$ could have significant consequences. Therefore, attackers in non-collaborative scenarios should be careful to evaluate the probability of correctly guessing $M^{*}$ against the added costs of failure -- either in terms of retaliation or of the strategic risks of being detected or identified by honest agents.

As a result of this discussion, for competitive scenarios involving SRA3P, we propose two variants of the classical attack, employed by attackers who are aware of the presence of active competitors. We term the variants \emph{strong} and \emph{competitive} attack, differing with respect to attacker knowledge. If the attacker is aware of the identity of the competitor, he will employ the strong attack, whereas he will resort to the competitive attack when only the competitor's presence is known.   
These new attack behaviors
are also oracle-type (transmission step, see~\cite{flaws} for a taxonomy of flaws and attacks) and are shown in Table~\ref{SRA3Pattacks}C and ~\ref{SRA3Pattacks}D.

The main difference between the two non-collaborative attack behaviors 
lies in the method of delivery of fake messages to $E_{1}$. If the competitor's identity is known, $E_{2}$ can ensure that the fake message is seen even if $E_{1}$ is not paying attention to $E_{2}$'s traffic: $E_{2}$ sends the fake message directly, using the network primitive \mathii{send}. 
If, on the other hand, $E_{1}$'s identity is unknown, $E_{2}$ is forced to rely on a reasonable prediction of $E_{1}$'s behavior and thus injects the fake message, impersonating $A$. The misleading message $M_{fake}$ is successfully delivered if $E_{2}$ is present in $E_{1}$'s dataset and $E_{1}$ spies it.
 If $E_{1}$ does not gain $M_{fake}$, $E_{2}$ fails to pollute the competitor's knowledge but does not compromise his own ability to observe $M^{*}$

SRA3P is such that all attentive attackers can potentially acquire the secret if an attack on the initiator $A$ is carried out. Exclusive knowledge of the secret can only occur through two mechanisms: through the outcome of erase requests (which is not under the control of network agents) or by misleading other attackers into interpreting a fake message as $M^{*}$.

An attack is successful if it goes undetected by the initiator $A$, who then transmits $M$ in the clear as $M^{*}$. Our agents are intelligent and they make use of all information available to perform in-protocol detection of attacks. With respect to SRA3P, a clear indication for $A$ consists in receiving a duplicate response from agents posing as $B$; 
under this circumstance, $A$ concludes that there has been a security
violation and halts the execution of the protocol to protect the secret $M$. 

From the attackers' perspective, an ongoing attack can be detected by observing that the message transiting on the network in step (2) is equal to the message  $\{\!| M |\!\}_{K_{A}}$ transiting on step (1). The attack trace is unambiguous to spying attackers. 
SRA3P is very unfriendly for attacker labeling: identifiers do not transit on the network, neither in the clear nor encrypted. Decisional processes cannot rely on any conclusive information concerning the identity of the agents involved in a given protocol run and must resort to inference on the basis of their current knowledge.

\subsection{Attacker configuration and outcomes of interaction}

We examine the outcome of attacks carried out in a non-collaborative environment in six cases, corresponding to different conditions of knowledge and belief for two attackers, $E_1$ and $E_2$. Refer to Table~\ref{SRA3Pcases} for a synthetic view of the message exchanges in each configuration.
In order to completely specify agent behavior, we state the following:

\begin{enumerate}
 \item An attacker who spies, before starting his own attack, the attack trace $\{\!| M |\!\}_{K_{A}}$ transiting on the network moves on to step (3) of his chosen attack (strong or competitive). If the attacker spies the attack trace after sending $\{\!| M |\!\}_{K_{A}}$ himself, then he requests that the message be erased. In our set-up, an erase-request always prevents the message from reaching its honest recipient, although other attackers cannot deterministically be prevented from spying it. This behavioral rule accounts for attackers being aware that duplicate messages can be exploited to perform attack detection. 
 \item An attacker that is employing a competitive attack (as he is aware of the presence of active competitors) continues spying on the network even after receiving the first message. 
 \item An attacker may learn that he has incorrectly classified an agent as honest. We do not wish to focus here on decisional processes for agent classification and therefore we posit that the decisional process is an oracle for the identifier of the mislabeled agent. We stipulate that, whenever evidence that an agent has been mislabeled arises, the decisional processes of the agents allow relabeling in \dishonest{} the dishonest agent who has triggered the anomalous situation detected. For completeness, we explicitly mention in case 1.T1-B which choices would be available to the agent, should the decisional process yield incorrect answers. 
 \item We posit that \cansee{} for $A$'s opening message comprises both $E_1$ and $E_2$; if this were not the case, only one attacker would be active in the run of the protocol examined.
 \item  We postulate that \cansee{} yields the entire attacker set for the message sent in step (3) by the honest agent $A$. If this were not the case, then only some (one) of the intruders could acquire $M^{*}$. For the sake of concisiveness, in the rest of this section we discuss explicitly only the situation posited. Refer to Section \ref{SRA3Pextended} for detailed analysis of how outcomes are affected by \cansee{}. 
\end{enumerate}
$ $\\
\noindent \textbf{Case 1: $E_{1}$ and $E_{2}$ know each other as honest.}\\
$E_{1}$ and $E_{2}$ know each other's identifiers (i.e.\ they are paying attention to each other: $E_{1} \in D_{E_{2}} \text{ and } E_{2} \in D_{E_{1}}$), but they are both mistaken in that they have labeled the other as honest ($E_{1} \in $ \honest{E_2} and $E_{2} \in$ \honest{E_1}).
Initially, $E_1$ and $E_2$ are unaware of active competitors and mount the classical attack 
The first between $E_{1}$ and $E_{2}$ to send the message at step (2) reveals to the other that he has incorrectly classified an agent. Without loss of generality, let us suppose that $E_{1}$ attacks first. $E_{2}$ employs his decisional processes to identify the mislabeled attacker. 
\\
\textbf{(1.T1-A)}: \emph{$E_{1}$ is identified as an attacker by $E_{2}$.}
$E_{2}$ switches to the strong attack, with the goal of gaining exclusive access to $M$. 
In step ($3_2$), $E_{2}$ sends a fake message to the unsuspecting competitor $E_{1}$, who is expecting a message from $A$ containing $M^{*}$ on the clear. $E_{1}$ may now think that he has successfully completed the attack, but in fact he did not acquire the secret $M$. After receiving $M_{fake}$, $E_{1}$ stops monitoring the network, according to the classical attack behavior. 

If $E_{1}$ continues to spy, he will also acquire $M^{*}.$ However, $E_{1}$ finds himself in a situation of uncertainty, as he is not able to determine if it is $M_{fake}$ or $M^{*}$ (or neither) that comes from $A$. $E_1$ can at most determine that there is an unlabeled active competitor, one that he has not previously identified in \dishonest{E_1}.
\\
\textbf{(1.T1-B)}: \emph{$E_{2}$ fails to identify $E_{1}$ as a dishonest agent.}
$E_{2}$ has two strategies available: i) risk revealing himself as an attacker and employ the strong attack against all agents he is attentive to (with the exception of the initiator of the protocol); ii) employ the competitive attack with partial impersonification.

$ $\\
\noindent \textbf{Case 2: $E_{1}$ and $E_{2}$ know each other as attackers.}\\
Both $E_{1}$ and $E_{2}$ (correctly) think that there are active competitors; they know the competitor's identity and thus both follow the strong attack. The attack trace prescribes waiting for a competitor to start the attack procedure, by sending $\{\!| M |\!\}_{K_{A}}$ to $A$. Both attackers are waiting for the other to take action. The situation could result in a deadlock, but the attackers know that a message has been erased and that $A$ is waiting for an answer.

The attackers wait for a reasonable amount of time and then one takes the initiative. Let us suppose that it is $E_{2}$ who first answers $A$. The strong attack consists in polluting the knowledge base of the competitor with a fake message. Both attackers send their fake messages ($M'$ and $M''$), thereby recreating the uncertainty of the previous case. This time the uncertainty spreads over both attackers and none dominates the other.


\begin{table}[ht]
\begin{center}
\scalebox{0.67}{
\begin{tabular}{cc}
\hline  
{\bf T1: cases 1, 4, 5, 6} & {\bf T2: case 2} \\
\hline \\
$
\begin{array}{rll}
(1\phantom{_{1}}) & A \to E_{1,2}(B)  & : \{\!| M |\!\}_{k_{A}} \\
(2_{1}) & E_{1}(B) \to A  & : \{\!| M |\!\}_{k_{A}} \\
(3_{2}) & E_{2}(A) \to E_{1}  & :  M_{fake} \\
(3\phantom{_{1}}) & A \to E_{[1],2}(B)  & :  M^{*} \\
\end{array}
$
&
$
\begin{array}{rll}
(1\phantom{_{1}}) & A \to E_{1,2}(B)  & : \{\!| M |\!\}_{k_{A}} \\
(2\phantom{_{1}}) & E_{2}(B) \to A  & : \{\!| M |\!\}_{k_{A}} \\
(3_{1}) & E_{1}(A) \to E_{2}  & :  M' \\
(3_{2}) & E_{2}(A) \to E_{1}  & :  M'' \\
(3\phantom{_{1}}) & A \to E_{1,2}(B)  & :  M^{*} \\
\end{array}
$
\\
&\\
\hline
{\bf T3: case 3} & {\bf T4: cases 4, 5 and 6} \\
\hline \\
$
\begin{array}{rll}
(1\phantom{_{1}}) & A \to E_{1,2}(B)  & : \{\!| M |\!\}_{k_{A}} \\
(2_{1}) & E_{1}(B) \to A  & : \{\!| M |\!\}_{k_{A}} \\
(2_{2}) & E_{2}(B) \to A  & : \{\!| M |\!\}_{k_{A}} \\
\end{array}
$
& 
$
\begin{array}{rll}
(1\phantom{_{1}}) & A \to E_{1,2}(B)  & : \{\!| M |\!\}_{k_{A}} \\
(2_{2}) & E_{2}(B) \to A  & : \{\!| M |\!\}_{k_{A}} \\
(2_{1}) & E_{1}(B) \to E_{2}(A)  & : \{\!| M |\!\}_{k_{A}} \\
(3_{2}) & E_{2}(A) \to E_{1}  & :  M_{fake} \\
(3\phantom{_{2}}) & A \to E_{1,[2]}(B)  & :  M^{*} \\
\end{array}
$
\\\\
\hline
\end{tabular}
}
\end{center}
\caption{
Traces for non-collaborative attacks against SRA3P. Traces are exhaustive aside for order of attackers.
Case 1: $E_1$ and $E_2$ know each other as honest. 
Case 2: $E_1$ and $E_2$ know each other as dishonest. 
Case 3: $E_1$ and $E_2$ are unaware of each other. 
Case 4: $E_2$ knows $E_1$ as honest. 
Case 5: $E_2$ knows $E_1$ as dishonest. 
Case 6: $E_2$ knows $E_1$ but has not yet established a belief on $E_1$'s honesty.
}
\label{SRA3Pcases}
\end{table}

$ $\\
\noindent \textbf{Case 3: $E_{1}$ and $E_{2}$ are unaware of each other.}\\
$E_{1}$ and $E_{2}$ are unaware of the other's presence -- i.e.\ they are not paying attention to the other's activity ($E_{1} \notin D_{E_{2}}$ and $E_{2} \notin D_{E_{1}}$). 
Thus, both $E_{1}$ and $E_{2}$ employ the classical attack. The attackers, not paying attention to the other's communications, do not realize that an attack trace is transiting on the network. $A$ receives a duplicate message, that he correctly interprets in terms of an ongoing attack. The attackers are detected, even if not explicitly identified. $A$ abandons the protocol to keep the secret $M$ safe.

$ $\\
\noindent \textbf{Case 4: $E_{2}$ knows $E_{1}$ as honest.}\\
$E_{2}$ is not aware of other attackers and can choose to attack right away or wait a reasonable time to try detecting a mislabeled attacker. 
\\
\textbf{(4.T1)}: \emph{$E_{2}$ waits and $E_{1}$ starts the classical attack.}
$E_{2}$ has the chance of detecting $E_{1}$ as an attacker and starts the strong attack. The situation is reduced to case 1. If $E_{1}$ continues to listen on the network after the end of his (unsuccessful) attack, he realizes that he is in a situation of uncertainty, not knowing which between $M^{*}$ and $M_{fake}$ is $A$'s secret. $E_{1}$ is now certain that an attacker is present but he doesn't know who, because the identifier $E_{2}$ is not in $E_{1}$'s proprietary dataset. 
$E_{1}$ can thus switch to an exploratory strategy, using the inflow-spy rule for the subsequent runs of the protocol. 
\\
\textbf{(4.T4)}: \emph{$E_{2}$ starts the classical attack.}
Not having $E_{2}$'s identifier in his dataset, $E_1$ does not pay attention to the message and does not notice the attack trace transiting. $E_{1}$ continues his attack and sends $\{\!| M |\!\}_{k_{A}}$. In step ($2_1$), $E_{2}$ detects the dishonesty of $E_{1}$ and switches to the strong attack. 
There is an important difference with respect to case 4.T1: $E_{2}$ erases the message sent by $E_{1}$ to $A$, thereby preventing $A$ from detecting a duplicate message and protecting his own attack.

$ $\\
\noindent \textbf{Case 5:  $E_{2}$ knows $E_{1}$ as dishonest.}\\
Only one out of the two attackers $E_{1}$ and $E_{2}$ is paying attention to the other and knows his identifier. Here we consider $E_{1} \in $ \dishonest{E_2} and $E_{2} \notin D_{E_{1}}$.
$E_{2}$ is sure of the presence of a competitor and knows his identifier. When $A$ initiates the protocol, $E_{2}$ waits for $E_{1}$ to start the attack and prepares to send a fake message in step ($3_2$), employing the strong attack (case 5.T1). If $E_{1}$ does not send $\{\!| M |\!\}_{k_{A}}$ within a reasonable time (case 5.T4), $E_{2}$ performs the attack in step ($2_{2}$). This message goes undetected by $E_{1}$, who will send his message at a later point. $E_{2}$ is aware that another attacker is present and is on the watch for a replicate attack message, which he erases. If $E_{1}$ acts first, the sequence of messages is the same as in case 1; otherwise, the sequence is the same as in case 4.T4. 

If $E_{1}$ continues to spy after receiving $M_{fake}$, he can realize that he is in uncertainty with respect to $M$ and can therefore deduce the presence of an unknown attacker. 
$E_{1}$ moves on to employing exploratory versions of the \mathii{spy}-rules to try gaining information about the identity of the competitor. 

$ $\\
\noindent \textbf{Case 6: $E_{2}$ knows $E_{1}$ but he is unsure of $E_{1}$'s honesty.}\\
Only one out of the two attackers $E_{1}$ and $E_{2}$ is paying attention to the other and knows his identifier. Here we consider $E_{1} \in $ \notknow{E_2} and $E_{2} \notin D_{E_{1}}$. 
This case reduces to cases 5.T1 and 5.T4, according to who first initiates the attack by sending $\{\!| M |\!\}_{k_{A}}$. In case 6.T1, $E_{1}$ opens the attack, whereas in case 6.T4 it is $E_{2}$ who opens. In all cases, $E_{2}$ has a clear advantage because he is paying attention to $E_{1}$'s messages but his own messages are not being attended to. In addition to what happens in case 5, $E_{2}$ has the opportunity to correctly label $E_{1}$: $E_2$ moves $E_1$'s identifier from \notknow{E_2} into \dishonest{E_2}.


\subsection{Success criteria for competitive attackers and honest agents}

Attackers in the SRA3P scenario have a complex success criterion. The best possible result for an attacker consists in i) violating security without the honest agent realizing it and ii) making it such that the other attackers conclude their attacks with false information ($M_{fake}$ taken for $M^*$) and without realizing that the information is false. This set of conditions describes an attacker with complete dominance over the system -- both over honest agents and over his competitors. As shown for SRA3P (and for BME in Section~\ref{sec:case-study}), in competitive scenarios with equal-opportunity attackers it is not possible, in general, to ensure a complete victory under all circumstances. 

The result of an attack depends on the strategy and on the knowledge conditions of all the active agents. As a consequence, a competitive agent will try to secure the best result (compatibly with his knowledge of the system) and he will strategically evaluate if it is preferable, for example, to risk being identified as an attacker by other agents or to increase the degree of uncertainty of the competitors. 
A competitive agent attacking SRA3P evaluates the following factors as part of his success criterion:
\begin{enumerate}
 \item Success in gaining the secret protected by the security system (or, more generally, in invalidating the target properties of the protocol). Because SRA3P is vulnerable to the classical attack, a single attacker without competition is always successful. The first priority of our competitive attackers is preserving the success of their own attacks, even in the presence of active competitors.
 \item Absence of uncertainty on the secret.
 \item Exclusivity in access to the secret.
 \item Effects on competitors: denying competitors either access to the secret or certainty on it. The ideal case for a competitive attacker is negating access to the secret and at the same time inducing competitors to think that they have succeeded. 
 \item Possibility of being identified as an attacker by other attackers. Attackers are aware that knowing the dishonesty of an agent is an advantage, therefore they seek to limit the situations in which they can be detected or identified through an explorative \mathii{spy}-rule. A good example of this strategy is the difference between the two non-collaborative attacks against SRA3P -- employing a direct \mathii{send} to the competitor or relying on the prediction that the competitor will spy.
 \item Possibility of being identified as an attacker by honest agents. 
 \item Possibility of identifying competitors -- and thus of acquiring a strategic advantage for later runs of the protocol.
\end{enumerate}
\noindent An honest agent that uses SRA3P distinguishes five relevant conditions, each associated to a different level of alarm:
\begin{enumerate}
 \item No attacker has gained the secret and the secret has correctly reached the intended recipient (security). Since SRA3P is vulnerable to attacks, in the presence of attackers this condition never occurs. 
 \item No attacker has succeeded in gaining the security secret, but the secret has not reached its intended recipient (stalemate, deadlock). For SRA3P, this condition occurs whenever the initiator detects duplicate messages before step (3), e.g.\ in case 3. 
 \item One or more attackers have gained the security secret but the honest agent has detected the attack (restart).
 \item One or more attackers have gained the security secret, the honest agent has detected the attack and has also acquired new knowledge on the identity of the attacker (retaliate and restart).
 \item One or more attackers have gained the security secret but the attack has not been detected (security failure). 
\end{enumerate}
During a protocol run, the proprietary datasets evolve in different ways according to the roles and the knowledge of the agents. The interpretation of messages -- and along with it the behavior -- can vary, both according to prior knowledge on the system and according to strategic considerations.

In Table~\ref{table:GuardianSRA3P}, we show the effects of introducing a guardian $G$ for SRA3P, configured as one of the competitive attackers described in the case study. Compared to the guardian for BME (see Section \ref{sec:defense}), a guardian for SRA3P appears to be less effective, in that it prevents $E$ from successfully carrying out his attack in fewer cases. However, it must be noted that SRA3P is a much harder protocol to defend because it does not entail that attacker success is mutually exclusive. Remarkably, $G$ can be effective even when he is not aware of $E$'s presence. The effectiveness of a guardian for SRA3P is comparable to the case of BME, if honest agents can detect and mount retaliatory attacks whenever attackers guess the wrong secret and use it to communicate with honest agents.

\begin{table}[ht]
\begin{center}
\scalebox{0.90}{
\begin{tabular}{c|cccc}
\hline
\footnotesize \!\!\!\!\!
\textbf{~~canSee~~}	 \!\!\!\!\! & 
\footnotesize \!\!\!\!\!
\textbf{~~Case 2~~}  \!\!\!\!\! & 
\footnotesize \!\!\!\!\!
\textbf{~~Case 3~~}  \!\!\!\!\!& 
\footnotesize \!\!\!\!\!
\textbf{~~Cases 1+4,5,6} ~  \!\!\!\!\!& 
\footnotesize \!\!\!\!\!
\textbf{~Cases 4,5,6} ~  \!\!\!\!\!\\
\footnotesize
	& 
\footnotesize
\textbf{~~} & 
\footnotesize
 & 
\footnotesize
\textbf{~~$E\in$\attentive{G}~} & 
\footnotesize
\textbf{~ $G\in$\attentive{E}~} \\
\hline
\puntello
\footnotesize $\{ E, G\}$ & \canHelp$^{+}$ & \helpAll & \canHelp & \\
\puntello
\footnotesize $\{ G\}$ & \helpAll & \helpAll & \helpAll & \helpAll \\
\puntello
\footnotesize $\{ E\}$ & \canHelp$^{+}$ & \helpAll & \canHelp & \\
\hline
\end{tabular}
}
\end{center}
\caption{
Effects of introducing a guardian $G$ for SRA3P when attacker $E$ is active. $G$ operates according to the same strategy as the attackers in the case study. $G$'s active interference results in $E$ failing to acquire the secret (\helpAll), in being sometimes uncertain (\canHelp) or in being always uncertain (\canHelp$^{+}$).
} 
\label{table:GuardianSRA3P}
\vspace{-0.5 cm}
\end{table}
%

\section{Extended tables for SRA3P}\label{SRA3Pextended}
In this appendix, we present a detailed view of the outcome of an attack
carried out against SRA3P and involving only the non-collaborative
attackers $E_1$ and $E_2$. Refer to Table~\ref{SRA3Pattacks} for a
definition of SRA3P and attacker behavior against SRA3P and to Table~\ref{SRA3Pcases} for attack traces and cases.

For each case, we report the following subcases (columns):
\begin{itemize}
 \item attacker $E_1$ is using the classical attack and stops spying on the network after receiving the first message that he can interpret as $M$.
 \item attacker $E_1$ continues to spy on the network even after receiving the first message that can be interpreted as $M$, with all possible values of the set \cansee{} for $A$'s response in step (3). If an attacker is not in canSee, he fails regardless of the number of fake messages dispatched.
\end{itemize}
For each attacker role, we describe:
\begin{itemize}
 \item  \textbf{(Attack)} which attack has been used (classical or strong) or if there has been a switch from the classical to the strong attack during the protocol run (Cl $\rightarrow$ Str). 
 \item \textbf{(Detection)} the ability to acquire further information on competitors. Possible values are: none performed (none); none possible, because the agent already has a correct understanding of the situation (none (c)); in-protocol detection, by spying the attack trace when no competitor is known ((in) trace); post-protocol detection, by realizing that more than one candidate $M$ has been spied and an unknown competitor is responsible for the uncertainty ((post) uncertainty) -- with the variant (post $\exists$) uncertainty to also signal that the identifier of the previously unknown competitor is not in \attentive{}.
 \item  \textbf{(Messages)} the set of messages that can be interpreted as $M$. $M!$ indicates that only $M$ has been spied; $M^+$ indicates that more than one message, including $M$, has been spied; $M_{fake}$ that only fake messages have been spied; none, to indicate that no message has been spied during the protocol run.
 \item  \textbf{(Result)} the result of the protocol run. Possible results are: full failure (the attacker does not acquire $M$ and takes a fake message for the secret), failure (the attacker does not acquire $M$ and realizes it), uncertainty (the attacker acquires the secret $M$ along with other fake messages), success (the attacker knows $M$ without uncertainty), dominance (the attacker succeeds and all his competitors fully fail).
\end{itemize}
For honest agents we show only the result: either security failure or attack detection through duplicate messages.

The last two rows in each table show the outcomes when a guardian $G$ is introduced along with a single (competitive) attacker $E$. $G_{E_1}$ plays the role of $E_1$ against $E$ playing $E_2$ and $G_{E_2}$ plays the role of $E_2$ against $E_1$. Similarly to attackers, we show for $G$ the possible conclusions that can be drawn on attacker identity and the actual security. Security can be: \emph{compromised}, if $E$ known $M$ with certainty; \emph{uncertain E}, if $E$ known $M$ but cannot identify it with certainty; \emph{restored}, if $E$ fails to acquire $M$; \emph{ enforced}, if thanks to $G$ being present, flags were raised for $A$ that allow $A$ to detect an ongoing attack and abort the protocol to protect $M$.

\begin{table}[ht]
\begin{center}
\scalebox{0.57}{
\newcommand{\mc}[3]{\multicolumn{#1}{#2}{#3}}
\begin{tabular}{|cc|c|c|c|c|c}
\hline
\textbf{Case 1} & & \textbf{$E_1$ stops} & \mc{3}{c|}{\textbf{$E_1$ continues and canSee($M^{*}$) = }} \\
\textbf{Agent} & \textbf{~Feature~}  & & $\{E_1, E_2\}$ & $\{E_1\}$ & $\{E_2\}$ \\
\hline
\hline
$E_{1}$ & Attack & Classical & Classical & Classical & Classical \\
 & Detection & none & (post) uncertainty  & (post) uncertainty & none \\
 & Messages & $M_{fake}$ & $M^{+}$ & $M^{+}$ & $M_{fake}$ \\
 & Result & full failure & uncertainty & uncertainty & full failure \\
\hline
$E_{2}$ & Attack & Cl $\rightarrow$ Str & Cl $\rightarrow$ Str  & Cl $\rightarrow$ Str  & Cl $\rightarrow$ Str  \\
 & Detection & (in) trace & (in) trace  & (in) trace  & (in) trace \\
 & Messages & $M!$ & $M!$ & none & $M!$ \\
 & Result & dominance & success & failure & dominance \\
\hline
$A$ & Result & failure & failure & failure & failure \\
\hline
${G_{E_1}}$ & Detection & none & (post) label & (post) label & none \\
 & Security & compromised & compromised & restored & compromised \\
\hline
$G{_{E_2}}$ & Detection & (in) label & (in) label & (in) label & (in) label \\
  & Security & restored & uncertain $E$  & uncertain $E$ & restored  \\
\hline
\hline
\end{tabular}
}
\end{center}
\caption[Results SRA3P, case 1 detail]{Overall SRA3P results, detailed view of \textbf{case 1}: $E_1$ and $E_2$ know each other as honest.
 If $E_1$ is no longer listening on the network, only $E_2$ can place an erase request in step (3) and thus can acquire the message $M^{*}$ with certainty. If the competitor $E_1$ continues to eavesdrop, the dominant intruder can fail to acquire $M^{*}$ whenever $E_2 \not\in$ canSee. If, on the other hand, it is the attacker at disadvantage ($E_1$) that is not in canSee, then $E_1$ fails regardless of the number of fake messages.}
\label{Table:SRA3Pcase1_extended}
\end{table}

\begin{table}[ht]
\begin{center}
\scalebox{0.67}{
\newcommand{\mc}[3]{\multicolumn{#1}{#2}{#3}}
\begin{tabular}{|cc|c|c|c|c|}
\hline
\textbf{Case 2}&  & \textbf{$E_1$ stops} & \mc{3}{c|}{\textbf{canSee($M^{*}$) = }} \\
\textbf{Agent} & \textbf{~Feature~}  & & $\{E_1, E_2\}$ & $\{E_1\}$ & $\{E_2\}$ \\
\hline
\hline
$E_{1}$ & Attack & -- & Strong & Strong & Strong \\
 & Detection & -- & none (c) & none (c) & none (c)\\
 & Messages & -- & $M^{+}$ & $M^{+}$ & $M_{fake}$ \\
 & Result & -- & uncertainty & uncertainty & full failure \\
\hline
$E_{2}$ & Attack & -- & Strong  & Strong  & Strong  \\
 & Detection & -- & none (c)  & none (c)  & none (c) \\
 & Messages & -- & $M^{+}$ & $M_{fake}$ & $M^{+}$ \\
 & Result & -- & uncertainty & full failure & uncertainty \\
\hline
$A$ & Result & -- & failure & failure & failure \\
\hline
${G_{E_1}}$ & Detection & -- & none (c) & none (c) & none (c)\\
 & Security & -- & uncertain $E$ & restored & uncertain $E$ \\
\hline
$G{_{E_2}}$ & Detection & -- & none (c) & none (c) & none (c) \\
  & Security & -- & uncertain $E$  & uncertain $E$ & restored  \\
\hline
\hline
\end{tabular}
}
\end{center}
\caption[Results SRA3P, case 2 detail]{Overall SRA3P results, detailed view of \textbf{case 2}:
$E_1$ and $E_2$ know each other as dishonest.}
\label{Table:SRA3Pcase2_extended}
\end{table}

\begin{table}[ht]
\begin{center}
\scalebox{0.57}{
\newcommand{\mc}[3]{\multicolumn{#1}{#2}{#3}}
\begin{tabular}{|cc|c|}
\hline
\textbf{Case 3}&  & \textbf{--} \\
\textbf{Agent} & \textbf{~Feature~}  & \\
\hline
\hline
$E_{1}$ & Attack & Classical \\
 & Detection & (post, $\exists$) failure \\
 & Messages & none \\
 & Result & failure \\
\hline
$E_{2}$ & Attack & Classical \\
 & Detection & (post, $\exists$) failure \\
 & Messages & none \\
 & Result & failure \\
\hline
$A$ & Result & detection (duplicates)\\
\hline
${G_{E_1}}$ & Detection & (post) $\exists$ \\
 & Security & enforced \\
\hline
$G{_{E_2}}$ & Detection & (post) $\exists$  \\
  & Security & enforced \\
\hline
\hline
\end{tabular}
}
\end{center}
\caption[Results SRA3P, case 3 detail]{Overall SRA3P results, detailed view of \textbf{case 3}:
$E_1$ and $E_2$ are unaware of each other. 
}
\label{Table:SRA3Pcase3_extended}
\end{table}

\begin{table}[ht]
\begin{center}
\scalebox{0.57}{
\newcommand{\mc}[3]{\multicolumn{#1}{#2}{#3}}
\begin{tabular}{|cc|c|c|c|c|}
\hline
& \mc{4}{c}{\textbf{ Case 4A: $E_1$ starts the attack}} &\\
\hline
& \mc{4}{c}{\textbf{ Case 4B: $E_2$ starts the attack}} &\\
 \hline
 &  & \textbf{$E_1$ stops} & \mc{3}{c|}{\textbf{canSee($M^{*}$) = }}  \\
\textbf{Agent} & \textbf{~Feature~}  & & $\{E_1, E_2\}$ & $\{E_1\}$ & $\{E_2\}$ \\
\hline
\hline
$E_{1}$ & Attack & Classical & Classical & Classical & Classical\\
 & Detection & none & (post $\exists$) uncertainty & (post $\exists$) uncertainty & none \\
 & Messages & $M_{fake}$ & $M^{+}$ & $M^{+}$ & $M_{fake}$ \\
 & Result & full failure & uncertainty & uncertainty & full failure \\
\hline
$E_{2}$ & Attack & Cl $\rightarrow$ Str & Cl $\rightarrow$ Str  & Cl $\rightarrow$ Str  & Cl $\rightarrow$ Str  \\
 & Detection & (in) trace & (in) trace  & (in) trace & (in) trace  \\
 & Messages & $M!$ & $M!$ & none & $M!$ \\
 & Result & dominance & success & failure & dominance \\
\hline
$A$ & Result & failure & failure & failure & failure \\
\hline
${G_{E_1}}$ & Detection & none & post ($\exists$) & post ($\exists$) & none\\
 & Security & compromised & compromised & restored & compromised \\
\hline
$G{_{E_2}}$ & Detection & (in) label & (in) label & (in) label & (in) label \\
  & Security & restored & uncertain $E$  & uncertain $E$ & restored  \\
\hline
\hline
\end{tabular}
}
\end{center}
\caption[Results SRA3P, case 4 detail]{Overall SRA3P results, detailed view of \textbf{case 4}:
$E_2$ knows $E_1$ as honest.}
\label{Table:SRA3Pcase4_extended}
\end{table}

\begin{table}[ht]
\begin{center}
\scalebox{0.57}{
\newcommand{\mc}[3]{\multicolumn{#1}{#2}{#3}}
\begin{tabular}{|cc|c|c|c|c|}
\hline
& \mc{4}{c}{\textbf{ Case 5A: $E_1$ starts the attack}} &\\
\hline
& \mc{4}{c}{\textbf{ Case 5B: $E_2$ starts the attack}} &\\
 \hline
 &  & \textbf{$E_1$ stops} & \mc{3}{c|}{\textbf{canSee($M^{*}$) = }}  \\
\textbf{Agent} & \textbf{~Feature~}  & & $\{E_1, E_2\}$ & $\{E_1\}$ & $\{E_2\}$ \\
\hline
\hline
$E_{1}$ & Attack & Classical & Classical & Classical & Classical\\
 & Detection & none & (post $\exists$) uncertainty & (post $\exists$) uncertainty & none \\
 & Messages & $M_{fake}$ & $M^{+}$ & $M^{+}$ & $M_{fake}$ \\
 & Result & full failure & uncertainty & uncertainty & full failure \\
\hline
$E_{2}$ & Attack & Strong & Strong & Strong & Strong \\
 & Detection & none (c)  & none (c)  & none (c) & none (c) \\
 & Messages & $M!$ & $M!$ & none & $M!$ \\
 & Result & dominance & success & failure & dominance \\
\hline
$A$ & Result & failure & failure & failure & failure \\
\hline
${G_{E_1}}$ & Detection & none & post ($\exists$) & post ($\exists$) & none\\
 & Security & compromised & compromised & restored & compromised \\
\hline
$G{_{E_2}}$ & Detection & none (c) & none (c) & none (c) & none (c) \\
  & Security & restored & uncertain $E$  & uncertain $E$ & restored  \\

\hline
\hline
\end{tabular}
}
\end{center}
\caption[Results SRA3P, case 5 detail]{Overall SRA3P results, detailed view of \textbf{case 5}:
$E_2$ knows $E_1$ as dishonest. }
\label{Table:SRA3Pcase5_extended}
\end{table}

\begin{table}[ht]
\begin{center}
\scalebox{0.57}{
\newcommand{\mc}[3]{\multicolumn{#1}{#2}{#3}}
\begin{tabular}{|cc|c|c|c|c|}
\hline
& \mc{4}{c}{\textbf{ Case 6A: $E_1$ starts the attack}} &\\
\hline
& \mc{4}{c}{\textbf{ Case 6B: $E_2$ starts the attack}} &\\
 \hline
 &  & \textbf{$E_1$ stops} & \mc{3}{c|}{\textbf{canSee($M^{*}$) = }}  \\
\textbf{Agent} & \textbf{~Feature~}  & & $\{E_1, E_2\}$ & $\{E_1\}$ & $\{E_2\}$ \\
\hline
\hline
$E_{1}$ & Attack & Classical & Classical & Classical & Classical\\
 & Detection & none & (post $\exists$) uncertainty & (post $\exists$) uncertainty & none \\
 & Messages & $M_{fake}$ & $M^{+}$ & $M^{+}$ & $M_{fake}$ \\
 & Result & full failure & uncertainty & uncertainty & full failure \\
\hline
$E_{2}$ & Attack & Strong & Strong & Strong & Strong \\
 & Detection & (in) label  & (in) label  & (in) label & (in) label \\
 & Messages & $M!$ & $M!$ & none & $M!$ \\
 & Result & dominance & success & failure & dominance \\
\hline
$A$ & Result & failure & failure & failure & failure \\
\hline
${G_{E_1}}$ & Detection & none & post ($\exists$) & post ($\exists$) & none\\
 & Security & compromised & compromised & restored & compromised \\
\hline
$G{_{E_2}}$ & Detection & (in) label & (in) label & (in) label & (in) label   \\
  & Security & restored & uncertain $E$  & uncertain $E$ & restored  \\

\hline
\hline
\end{tabular}
}
\end{center}
\caption[Results SRA3P, case 6 detail]{Overall SRA3P results, detailed view of \textbf{case 6}: 
$E_2$ knows $E_1$ but has not yet established a belief on $E_1$'s honesty.}
\label{Table:SRA3Pcase6_extended}
\end{table}

\end{document}